\newif\ifCVPR \CVPRtrue
\newif\ifdraft \draftfalse
\newif\ifanonymous \anonymousfalse
\newif\ifappendix \appendixtrue
\newif\ifarxiv \arxivtrue

\documentclass[10pt,twocolumn,letterpaper]{article}

\ifanonymous
\usepackage[review]{CVPR/cvpr}      %
\else
    \ifarxiv
        \usepackage[pagenumbers]{CVPR/cvpr} %
    \else
        \usepackage{cvpr}              %
    \fi
\fi

\usepackage{graphicx}
\usepackage{amsmath}
\usepackage{amssymb}
\usepackage{booktabs}

\usepackage[pagebackref,breaklinks,colorlinks,pdfencoding=auto]{hyperref}

\usepackage[capitalize]{cleveref}
\crefname{section}{Sec.}{Secs.}
\Crefname{section}{Section}{Sections}
\Crefname{table}{Table}{Tables}
\crefname{table}{Tab.}{Tabs.}

\def\cvprPaperID{6503} %
\def\confName{CVPR}
\def\confYear{2023}

\newif\ifCVPR \CVPRtrue
\newif\ifdraft \draftfalse
\newif\ifanonymous \anonymousfalse
\newif\ifappendix \appendixtrue
\newif\ifarxiv \arxivtrue

\usepackage{stackengine} %
\ifundef{\ifAAAI}      {\newif\ifAAAI \AAAIfalse}           {}
\ifundef{\ifanonymous} {\newif\ifanonymous \anonymoustrue} {}
\ifundef{\ifdraft}     {\newif\ifdraft \draftfalse}         {}
\ifundef{\ifarxiv}     {\newif\ifarxiv \arxivfalse}         {}

\ifAAAI
\else %
\usepackage{float} 
\usepackage[normalem]{ulem} %
\usepackage{hyperref} 
\usepackage{supertabular} 
\usepackage{wrapfig}
\usepackage[ruled,vlined,linesnumbered]{algorithm2e} %
\fi
\usepackage{comment}
\usepackage{graphicx}
\usepackage{amsmath}
\usepackage{xcolor}
\usepackage{enumitem}
\usepackage{epsfig}
\usepackage{xspace} %
\usepackage{url}
\usepackage{graphics}
\usepackage{multirow}
\usepackage{booktabs}
\usepackage{soul}
\usepackage{makecell}
\usepackage{cleveref}
\usepackage{dsfont}
\usepackage{diagbox}

\usepackage{xr}
\makeatletter
\ifundef{\addFileDependency} 
{
    \newcommand*{\addFileDependency}[1]{%
      \typeout{(#1)}
      \@addtofilelist{#1}
      \IfFileExists{#1}{}{\typeout{No file #1.}}
    }
}
{}
\makeatother

\ifundef{\myexternaldocument} 
{
    \newcommand*{\myexternaldocument}[1]{%
        \externaldocument{#1}%
        \addFileDependency{#1.tex}%
        \addFileDependency{#1.aux}%
    }
}
{}

\makeatletter
\ifundef{\onedot} 
{
\DeclareRobustCommand\onedot{\futurelet\@let@token\@onedot}
\def\@onedot{\ifx\@let@token.\else.\null\fi\xspace}
}
{}

\ifundef{\eg} {\def\eg{\emph{e.g}\onedot}} {} \ifundef{\Eg} {\def\Eg{\emph{E.g}\onedot}} {} 
\ifundef{\ie} {\def\ie{\emph{i.e}\onedot}} {} \ifundef{\Ie} {\def\Ie{\emph{I.e}\onedot}} {} 
\ifundef{\cf} {\def\cf{\emph{cf}\onedot}} {} \ifundef{\Cf} {\def\Cf{\emph{Cf}\onedot}} {} 
\ifundef{\etc} {\def\etc{\emph{etc}\onedot}} {} \ifundef{\vs} {\def\vs{\emph{vs}\onedot}} {} 
\ifundef{\wrt} {\def\wrt{w.r.t\onedot}} {} \ifundef{\dof} {\def\dof{d.o.f\onedot}} {} 
\ifundef{\iid} {\def\iid{i.i.d\onedot}} {}  \ifundef{\wolog} {\def\wolog{w.l.o.g\onedot}} {} 
\ifundef{\etal} {\def\etal{\emph{et al}\onedot}} {} 
\ifundef{\sup} {\def\sup{sup. mat\onedot}} {} 
\makeatother

\ifAAAI
\usepackage{amssymb}%
\fi
\usepackage{pifont}%

\newcommand{\smallmath}{\fontsize{8pt}{5pt}\selectfont}
\newcommand{\smallmathtext}[1]{{\fontsize{8pt}{20pt}\selectfont $#1$}}
\newcommand{\fixwithphantom}{\vphantom{$\bR^T$}}
\ifundef{\shortcite} {\newcommand{\shortcite}{\cite}} {}

\newcommand{\bF}{{\bf F}}

\newcommand{\bI}{{\bf I}}

\newcommand{\bP}{{\bf P}}

\newcommand{\bR}{{\bf R}}

\newcommand{\bV}{{\bf V}}

\newcommand{\loc}{\bP}
\newcommand{\velo}{\bV}

\newcommand{\mll}{\mathcal{L}}
\newcommand{\mm}{\mathcal{M}}
\newcommand{\nn}{\mathcal{N}}

\newcommand{\rr}{\mathcal{R}}
\newcommand{\mss}{\mathcal{S}}
\newcommand{\ww}{\mathcal{W}}
\newcommand{\zz}{\mathcal{Z}}
\newcommand{\Motion}{\mm}
\newcommand{\Loss}{\mll}

\newcommand{\R}{\mathds{R}} %
\newcommand{\E}{\mathds{E}} %

\newcommand{\mean}{\bar}
\newcommand{\norm}[1]{\left\lVert#1\right\rVert}

\ifdraft
\newcommand{\todo}[1]{{\color{red} #1}}
\newcommand{\sr}[1]{{\color{violet} #1}}
\newcommand{\srr}[1]{{\color{red}\textbf{SR:} #1}}
\newcommand{\srv}[1]{{\color{violet}\textbf{SR:} #1}}

\newcommand{\dcc}[1]{{\color{red}\textbf{DC:} #1}}

\newcommand{\kac}[1]{{\color{cyan}\textbf{KA:} #1}}

\newcommand{\plc}[1]{{\color{orange}\textbf{PL:} #1}}

\newcommand{\ilreplace}[2]{\pl{#1} \il{#2}}
\else
\newcommand{\todo}[1]{}
\newcommand{\sr}[1]{#1}
\newcommand{\srr}[1]{}
\newcommand{\srv}[1]{}

\newcommand{\dcc}[1]{}

\newcommand{\kac}[1]{}

\newcommand{\plc}[1]{}

\newcommand{\ilreplace}[2]{#2}
\fi

\newcommand{\algoname}{{MoDi}} %
\newcommand{\filtername}{{convolutional scaler}} 

\newcommand{\narroweq}{\!=\!}

\newcommand{\narrowmin}{\!-\!}
\newcommand{\narrowtimes}{\!\times\!}

\makeatletter
\newcommand{\paragraphtinyvert}
{
  \@startsection{paragraph}{4}
  {\z@}{2pt plus -0pt minus 0pt}{-0.5em}
  {\normalsize\bf}
}
\newcommand{\paragraphnovert}{%
  \@startsection{paragraph}{4}%
  {\z@}{0ex \@plus 0ex \@minus 0ex}{-0.5 em}%
  {\normalsize\bf}%
}
\newcommand{\subparagraphnovert}{%
  \@startsection{subparagraph}{5}%
  {3ex}{0ex \@plus 0ex \@minus 0ex}{-1em}%
  {\normalfont\normalsize\bfseries}%
}
\makeatother

\begin{document}

\title{\algoname{}: Unconditional Motion Synthesis from Diverse Data}

\iftrue
\author {
    Sigal Raab \textsuperscript{\rm 1} \quad
    Inbal Leibovitch \textsuperscript{\rm 1} \quad
    Peizhuo Li \textsuperscript{\rm 2} \\
    Kfir Aberman \textsuperscript{\rm 3} \quad
    Olga Sorkine-Hornung \textsuperscript{\rm 2} \quad
    Daniel Cohen-Or \textsuperscript{\rm 1} \\ \\
\textsuperscript{\rm 1} Tel-Aviv University \quad
\textsuperscript{\rm 2} ETH Zurich \quad
\textsuperscript{\rm 3} Google Research \\
{\tt\small sigal.raab@gmail.com}
}
\fi

\ifCVPR
\ifCVPR
    \twocolumn[{%
    \renewcommand\twocolumn[1][]{#1}%
        \maketitle
        \begin{center}
        \captionsetup{type=figure}
        \includegraphics[width=\textwidth]{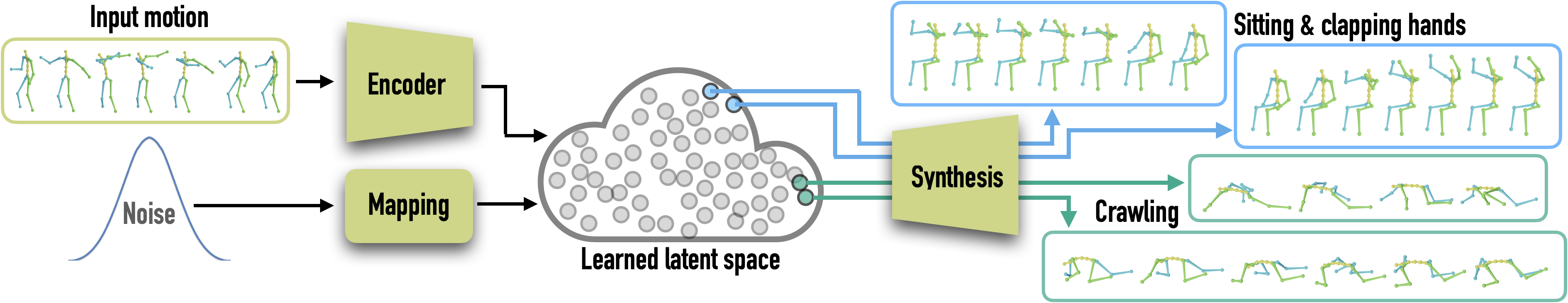}
        \captionof{figure}{Our generative model is learned in an unsupervised setting from a diverse, unstructured 
        and unlabeled motion dataset and yields a highly semantic, clustered, latent space that facilitates synthesis operations. 
        \sr{An encoder and a mapping network enable the employment of real and generated motions, respectively}.
        }
        \end{center}%
    }]

\else

\begin{figure}[t]

\centering
\includegraphics[width=\linewidth]{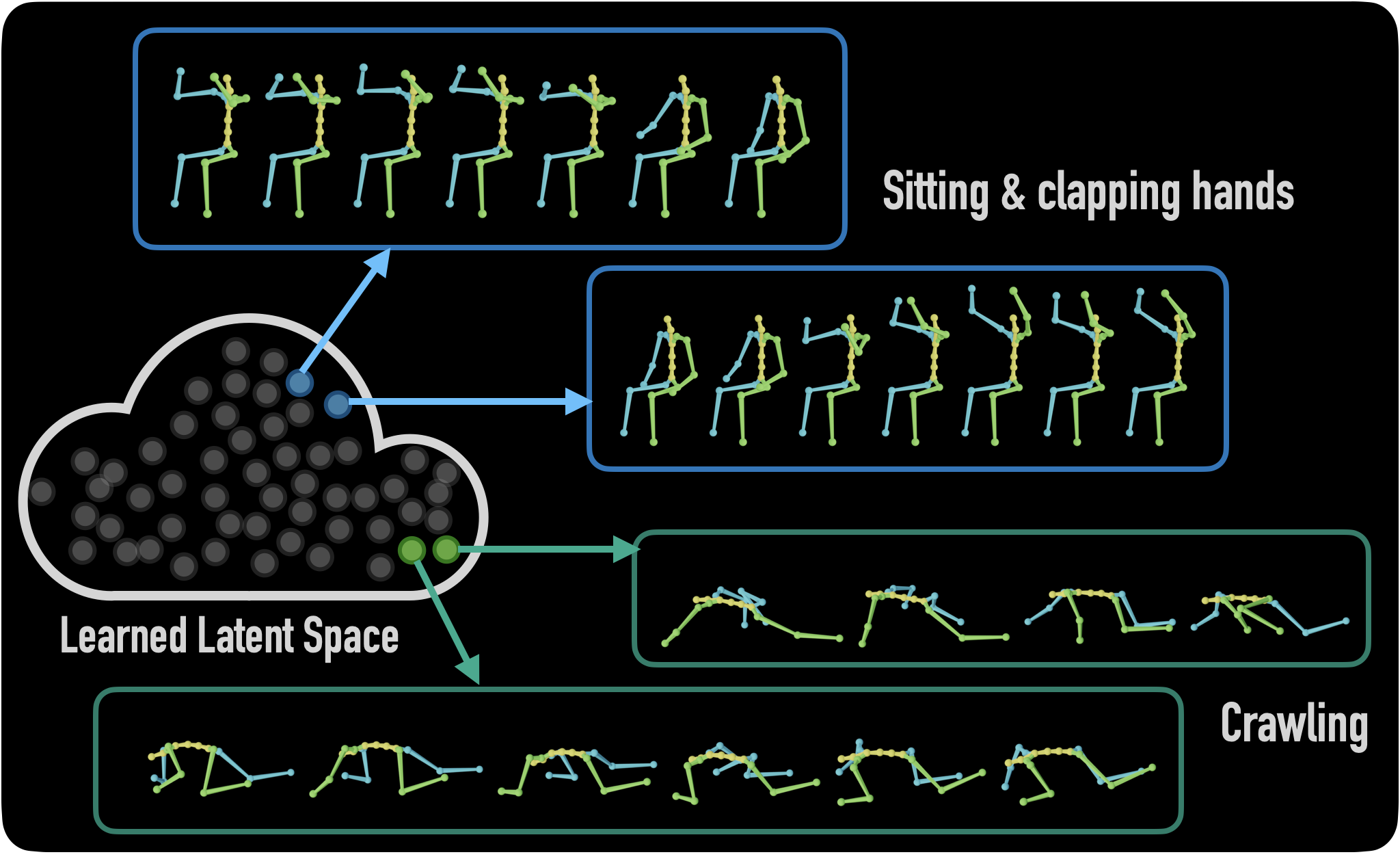}

\caption{Our generative model is learned in an unsupervised setting from a diverse, unstructured and unlabeled motion dataset and yields a highly semantic, clustered, latent space that facilitates \sr{various applications}.
Note that each single latent code represents a whole motion sequence. 
}
\label{fig:teaser}

\end{figure}

\fi
\fi 
\maketitle

\begin{abstract}
\vspace{-5pt}

The emergence of neural networks has revolutionized the field of motion synthesis. Yet, learning to unconditionally synthesize motions from a given distribution remains challenging, especially when the motions are highly diverse.
In this work, we present \algoname{} -- a generative model trained in an unsupervised setting from an extremely diverse, unstructured and unlabeled   dataset. 
During inference, MoDi can synthesize high-quality, diverse motions. 
Despite the lack of any structure in the dataset, our model yields a well-behaved and highly structured latent space, which can be semantically clustered, constituting a strong motion prior that facilitates various applications including semantic editing and crowd simulation.
In addition, we present an encoder that inverts real motions into MoDi's natural motion manifold, issuing solutions to various ill-posed challenges such as  completion from prefix and  spatial editing.
Our qualitative and quantitative experiments achieve state-of-the-art results that outperform recent SOTA techniques.
Code and trained models 
\ifanonymous{will be released.}\else{are available at https://sigal-raab.github.io/MoDi.}\fi
\end{abstract}

\vspace{-5pt}
\section{Introduction}
\vspace{-5pt}

The field of motion synthesis includes a wide range of long-standing tasks 
whose goal is to generate a sequence of temporally coherent poses that satisfy given cues and/or spatio-temporal constraints and importantly, look natural. In particular, learning to synthesize human motion from a given data distribution is a challenging task, especially when the dataset is highly diverse, unstructured and unlabeled. 
In recent years, deep neural networks have become a popular tool for motion generation, and their excellent performance is imputed to their ability to learn motion priors from large scale datasets. However, learning a motion prior from a diverse dataset 
remains a challenge.

Previous works focus on synthesizing specific types of motion of limited diversity \cite{holden2017phase}, conditioned by a set of frames \cite{zhou2018auto} or by a label indicating a text or an action \cite{tevet2022human}.

In this work, we present \algoname{}, an unconditional generative model that synthesizes diverse motions.
\algoname{} is unsupervised and is trained on diverse, unstructured and unlabeled datasets, yielding a well-behaved, highly semantic latent space, facilitating a variety of synthesis operations. 

Our design is inspired by the powerful architecture of StyleGAN~\cite{karras2020analyzing}, which has become a foundation for synthesis in the imaging domain, as it learns a well-structured latent space that allows incredible semantic editing capabilities~\cite{bermano2022state}. 
However, there is a significant gap between the imaging and motion domains; Images possess a regularized 2D spatial structure with a relatively large number of degrees of freedom (DoF), while motion data is irregular, consisting of a skeletal graph with a temporal axis that has a smaller number of DoF.
To mitigate this gap, we have conducted a thorough study of potential operators (2D vs. 3D convolutions, with and without skeleton-aware \cite{aberman2020skeleton}), architectural variations (order and number of blocks, resolution, layers, etc.), and even proposed a new building block \sr{(a \filtername{})} that enables us to achieve state-of-the-art results in unconditional motion synthesis.

Our results show that \algoname{} learns a structured latent space that can be clustered into regions of semantically similar motions without any supervision. This latent space facilitates applications on diverse motions, including semantic editing, semantic interpolation between motions, and crowd simulation, that we show in the paper.

In addition, we present an encoder architecture that leverages the knowledge we acquired on the generative model architecture, to invert unseen motions into MoDi's latent space, facilitating the usage of our generative prior on real world motions. Inversion by an encoder enables us to project motions into the latent space within a feed forward pass, instead of optimizing the latent code which requires several minutes for a single input \cite{richardson2021encoding}. Importantly, an encoder can better project a given motion to the well-behaved part of the latent space and can better assists in solving ill-posed problems where only part of the data is presented in the input (e.g., motion prediction from prefix) as it learns to target its projections to the healthy regions of the latent space instead of overfitting to the partial data.

We evaluate our model qualitatively and quantitatively on the Mixamo~\cite{mixamo} and  HumanAct12~\cite{guo2020action2motion} datasets, and show that it outperforms SOTA methods in similar settings. 
The strength of our generative prior is demonstrated through the various applications we show in this work, \ilreplace{including}{ such as} semantic editing, crowed simulation, prediction from prefix, motion fusion, denoising, and spatial editing.

\section{Related Work} \label{sec:related_work}
\vspace{-5pt}

The emergence of neural networks has transformed the field of motion synthesis, and many novel neural models have been developed in recent years ~\cite{holden2015learning, holden2016deep}. Most of these models focus on specific human motion related tasks, conditioned on some limiting factors, such as motion prefix ~\cite{aksan2019structured,barsoum2018hp,habibie2017recurrent,yuan2020dlow,zhang2021we,hernandez2019human}, in-betweening \cite{harvey2020robust,duan2021single,kaufmann2020convolutional,harvey2018recurrent}, motion retargeting or style transfer~\cite{holden2017fast,villegas2018neural,aberman2019learning, aberman2020skeleton,aberman2020unpaired}, music ~\cite{aristidou2021rhythm,sun2020deepdance,li2021learn,lee2018listen}, action~\cite{petrovich2021action,guo2020action2motion,cervantes2022implicit, tevet2022human}, or text~\cite{tevet2022motionclip,zhang2021write, Petrovich2022temos,ahuja2019language2pose,guo2022generating,bhattacharya2021text2gestures, tevet2022human}. 

A large number of models focus on action conditioned generation. These works are closer in spirit to ours, hence in the following we elaborate about them. These models can be roughly divided to autoregressive~\cite{petrovich2021action,
fragkiadaki2015recurrent,zhou2018auto,maheshwari2022mugl,guo2020action2motion,habibie2017recurrent,Jang2020constructing,ghorbani2020b}, diffusion-based~\cite{tevet2022human} and GAN-based ~\cite{degardin2022generative, wang2020learning, yan2019convolutional,yu2020structure}.

Petrovich \etal \shortcite{petrovich2021action} 
learn an action-aware latent representation by training a VAE. They sample from the learned latent space and query a series of positional encodings to synthesize 
motion sequences conditioned on an action. They employ a transformer for encoding and decoding a sequence of parametric SMPL human body models.
Maheshwari \etal \shortcite{maheshwari2022mugl} 
generate single or multi-person pose-based action sequences with locomotion. They present
generations conditioned by 120 action categories. They use a Conditional Gaussian Mixture Variational Autoencoder 
to enable intra and inter-category diversity.
Wang \etal \shortcite{wang2020adversarial} employ a sequence of recurrent autoencoders. They replace the KL divergence loss by a discriminator to ensure the bottle neck distribution. 

Some GAN-based models are combined with factors that limit their generalization, such as Gaussian processes~\cite{yan2019convolutional} or auto encoders~\cite{wang2020learning,yu2020structure}.
Degardin \etal \shortcite{degardin2022generative} fuse the architectures of GANs and GCNs to synthesise the kinetics of the human body. Like us, they borrow a mapping network from StyleGAN~\cite{
karras2020analyzing}. However, their model does not utilize important aspects of StyleGAN such as multi-level style injection. As we demonstrate, these aspects significantly ameliorate the quality of the synthesized motions. Unlike the above conditional models, we present an unconditional method. 

Only a few works enable pure unconditioned synthesis.
Holden \etal \shortcite{holden2016deep} presented 
a pioneering work in deep motion synthesis.
Their latent space is not sufficiently disentangled, so they train a separated feed forward network for each editing task, while \algoname{} either performs editing in the latent space with no need to train an additional network (\cref{sec:latent_space}), or uses a single encoder for a variety of applications (\cref{sec:encoder_apps}). 
Another model that supports an unconstrained setting is MDM~\cite{tevet2022human}. Although they use state-of-the-art diffusion models, we demonstrate that \algoname{} outperforms their unconditional synthesis setting (\cref{sec:quantitative}).

In order to process motion in a deep learning framework, many existing works convert the motion into a pseudo image, where the joints and time-frames are equivalent to image height and width, and joint features (e.g., coordinates) are equivalent to RGB channels~\cite{holden2016deep,maheshwari2022mugl,hernandez2019human, petrovich2021action}. 
While this approach is straightforward and intuitive, joints are fundamentally different from image pixels in that they are not necessarily adjacent to each other as pixels are. 
A partial solution for this problem is presented in Tree Structure Skeleton Image (TSSI)~\cite{yang2018action}, where some of the joints are replicated to ensure skeletal continuity in convolution. However, TSSI cannot reflect all neighborhood degrees.

The emergence  of Graph-based convolutional networks has been adopted by the motion research community~\cite{yan2019convolutional}, since the human skeleton is naturally represented by a graph, where the joints and bones are represented with vertices and edges, respectively. A full motion is then considered as a spatio-temporal graph~\cite{yu2020structure, degardin2022generative}. Since a single kernel shared by all joints cannot capture the fine nuances of each joint, more advanced techniques ~\cite{aberman2020skeleton,yan2019convolutional} exploit the advantage of using finite size skeletons with predefined topology. Each skeletal joint is unique in the way it relates to its neighbors.
In our work, we adopt this approach and dedicate a unique kernel for each joint.

\section{Method} \label{sec:model}
\vspace{-5pt}

At the crux of our approach lays a deep generative model trained in an unsupervised manner on an extremely diverse, unstructured and unlabeled motion dataset. 
Our network receives a noise vector drawn from an i.i.d Gaussian distribution and outputs a natural, temporally coherent human motion sequence.
Once the generator is trained, the learned prior can be leveraged for various applications, and can be applied to either synthetic or real motions using an encoder model that receives a motion tensor and inverts it into the latent space of the generative model. 

In recent years, generative works in the image domain have attained unprecedented synthesis quality~\cite{brock2018large, kingma2018glow, ho2020denoising}, and our framework is inspired by one of the prominent methods -- StyleGAN~\cite{karras2020analyzing,karras2021alias}. However, StyleGAN as is \emph{cannot} be used for motion synthesis since there is a significant domain gap between images and motions that makes the adaptation non-trivial. First, images possess a regularized spatial structure with an inductive bias of pixel neighborhood which is strongly exploited, while motions are irregular, consisting of joints whose features are adjacent in a tensor but are unnecessarily adjacent in the skeletal topology.
Second, images have a relatively larger number of DoF comparing to the DoF of motion which is limited by the number of joints.

In order to bridge the gap, our architectural design employs structure-aware neural filters that enable us to cope with the irregular motion representation. Unlike previous works in the domain, we use 3D convolutions rather than 1D or 2D ones, facilitating essential modulation operators with a dedicated kernel for each skeletal joint. The benefit of 3D filters is detailed in \ifappendix{\cref{sec:neural_modules}. }\else{the sup. mat. }\fi
In addition, to compensate for the low number of DoF and prevent over-fitting, we engage a hierarchy that is shallower than the one used in the imaging domain. Moreover, we suggest a novel skeleton-aware convolutional-pooling filter to boost the performance of our networks.

Next, we discuss our structure-aware modules and network architecture, and present an inversion technique that projects a given motion into the learned latent space. In \cref{sec:latent_space} we show that our latent space is semantically clustered and demonstrate semantic editing applications, and in \cref{sec:encoder_apps} we demonstrate that ill-posed problems can be solved with our encoder. 
Finally, we show quantitative and qualitative evaluation of our framework and compare it to  state-of-the-art alternatives (Section~\ref{sec:experiments}).
We refer the reader to the supplementary video to see the results of our work.

\subsection{Motion Representation}
\vspace{-5pt}

We describe a motion using temporally coherent 3D joint rotations,
\smallmathtext{\bR \! \in \! \R^{T \times J \times K}}, where $T$, $J$ and $K$ are the numbers of frames, joints and rotation features, respectively. We found that unit quaternions (4D) attain the best empirical results when used for rotation representation.
The root joint position is represented by a sequence of global displacements, \smallmathtext{\loc \! \in \! \R^{T \times 3}}, and their velocities, \smallmathtext{\velo \! \in \! \R^{T \times 3}}.
In addition,\fixwithphantom{}  our network learns to refrain from foot sliding artifacts using binary foot contact labels,
\smallmathtext{\bF \! \in \! \{0,1\}^{T \times 2}}, that are concatenated to the joints axis. %
 We zero-pad the feature dimension of the root location 
and the foot contact labels to the size of the rotation feature, $K$, \fixwithphantom and add an extra dimension, so all entities ($\bR$, $\velo$ and $\bF$) \fixwithphantom possess the same number of features.
Altogether we have 
\smallmathtext{\bR \! \in \! \R^{T \times J \times K}} (unchanged), \smallmathtext{\hat{\velo} \! \in \! \R^{T \times 1 \times K}}, and \smallmathtext{\hat{\bF} \! \in \! \R^{T \times 2 \times K}}.
Once all features share the same size, we concatenate them and obtain the full motion space by  
{\ifAAAI \smallmath \fi
\begin{equation}
\vspace{-5pt}
\Motion_{full}\equiv\R^{T \times E\times K},
\end{equation}
} where $E \narroweq J \! +  \! 3$ is the number of entities ($\bR$, $\velo$ and $\bF$).

Let $\Motion_{nat}$ denote the space of natural motions that 
are plausible for humans to enact.
Each motion $m\in \Motion_{nat}$ is represented by a tuple, $[\bR_m,\hat{\velo}_m,\hat{\bF}_m]$.
Note that the sub-space of all human motions, $\Motion_{nat} \! \subset \! \Motion_{full}$, is extremely sparse, as most of the values in $\Motion_{full}$ correspond to unnatural or impossible human motions.

Our network has a hierarchical structure in which the represented motion is evolving from coarse motion representation to a finer one. 
At each level $\ell$ the number of frames, joints, entities and features is denoted by $T_\ell$, $J_\ell$, $E_\ell$ and $K_\ell$, respectively. 
The number of frames $T_\ell$ increases between two consecutive levels 
by a factor of 2, and the number of joints increases by a topologically specified 
factor in order to obtain a meaningful refinement of the skeleton~\cite{aberman2020skeleton,degardin2022generative}.
More representation considerations are detailed in \ifappendix{ \cref{sec:motion_rep}.}\else{the sup. mat. }\fi

\subsection{Structure-aware Neural Modules}\label{sec:operators}
\vspace{-5pt}
\begin{figure}
\centering
\includegraphics[width=.8\linewidth]{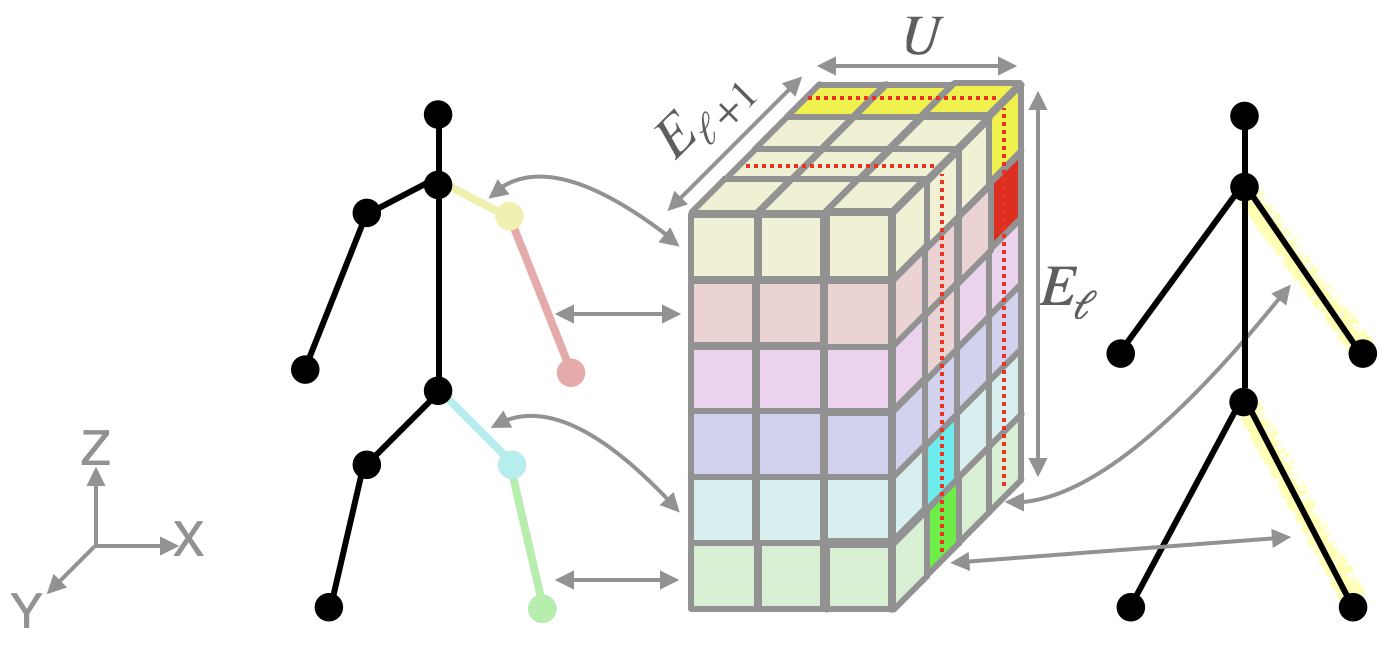} 

\setlength{\abovecaptionskip}{0pt plus 3pt minus 2pt}
\setlength{\belowcaptionskip}{-10pt plus 3pt minus 2pt}

\caption{
\sr{3D \filtername{}:
Each horizontal slice affects one entity in the fine character (left), and each vertical slice (xz plane) affects one entity in the coarse character (right). 
Each entity in the coarse character ``sees'' only weights related to relevant entities of the fine character, emphasised with saturated colors in the filter. 
}}
\label{fig:conv_pool}
\end{figure}

We consider the human skeleton as a directed graph, where the joints stand for vertices and the bones stand for directed edges.
We associate each skeletal joint with the edge that is directed towards it, hence they share the same features.
The root joint, to which no edge is directed, is associated with an abstract edge that starts at the origin. 

Some works~\cite{degardin2022generative,yu2020structure} use Graph Convolutional Networks (GCNs) for neural computation. Like GCNs, they employ the same kernels to all graph vertices. Unlike general graphs, the topology of the  skeleton is known in advance, and has a finite size. These facts can be exploited to get better sensitivity to each joint's unique role in the skeleton. We follow the works that exploit the knowledge of skeletal topology~\cite{yan2019convolutional,aberman2020skeleton} and dedicate separate kernels for each joint.

\sr{However, 
these works use naive pooling to up/down sample the skeletal (spatial) domain,
which are essentially mere copying and averaging.
Alternatively, we present a spatio-temporal convolutional operator, that scales the skeleton topology, as well as the temporal dimension.
We use a convolution filter during down sampling and transposed convolution for up sampling, both share the same filter architecture. We achieve the desired functionality by adding a dimension to the kernels for the out going joints, similar to the way a dimension is added for the out going channels.
The dimensions of each filter are then $K_{\ell+1} \narrowtimes K_\ell \narrowtimes E_{\ell+1} \narrowtimes E_\ell \narrowtimes U $, where $U$ is the filter width.
\cref{fig:conv_pool} visualizes our novel \filtername{} filter and \ifappendix{\cref{sec:neural_modules} }\else{the sup. mat. }\fi elaborates about it.

In addition, we use one existing skeleton-aware module, namely in-place convolution~\cite{aberman2020skeleton}, and add a third dimension to it too. 
The motivation for the extra dimension is the convenience of applying modulation, explained in \ifappendix{\cref{sec:neural_modules}. }\else{the sup. mat. }\fi  
\ifappendix{\cref{sec:neural_modules} }\else{The sup. mat. }\fi also describes skeleton-aware modules in existing works (convolutional and pooling).
}

\subsection{Generative Network Architecture} \label{sec:architecture}
\vspace{-5pt}

\begin{figure*}
\centering
\includegraphics[width=.9\linewidth]{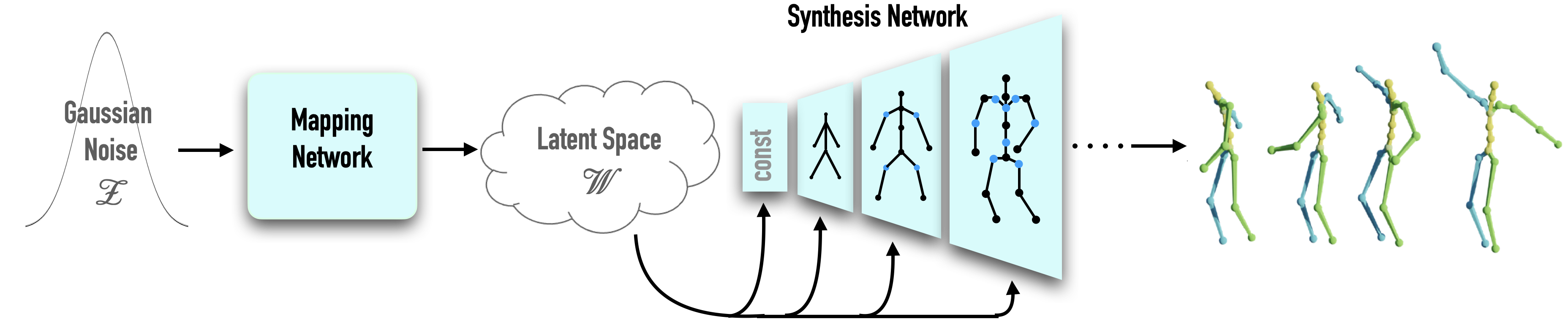} 

\setlength{\abovecaptionskip}{0pt plus 3pt minus 2pt}
\setlength{\belowcaptionskip}{-10pt plus 3pt minus 2pt}
\caption{Our motion generator combines structure-aware neural modules with a mapping network and style-codes injected to multiple levels of the generator. A detailed description of the architecture (e.g., layers, hyperparameters) is given in the \ifappendix{appendix.}\else{sup. mat.}\fi}
\label{fig:architecture}

\end{figure*}

Our network receives a noise vector drawn from an i.i.d Gaussian distribution, $\zz$, and outputs a natural, temporally coherent, human motion sequence, as depicted in \cref{fig:architecture}. Our generator $G$ consists of two main parts: a mapping network that maps noise into a well-behaved, structured, latent space, and a fully convolutional neural network that maps a learned constant and the latent code into the final motion.

\paragraphtinyvert{Mapping network}
Let $\zz= \nn(\vec{0},\bI)$ be a multivariate normal distribution.
Given a latent code $z\in\zz$, a non-linear mapping network produces a latent value, $w\in\ww$, where $\ww$ is known to be disentangled and well behaved, as studied for images~\cite{shen2020interfacegan,nitzan2021large} and for motions~\shortcite{degardin2022generative}.

\paragraphtinyvert{Motion synthesis network}

We use a hierarchical framework that learns to convert a learned constant into a motion representation via a series of skeleton-aware convolutional layers (\cref{sec:operators}), where the traditional skeletal pooling layer is replaced by our \filtername{}. The layers in the motion synthesis network are modulated by style-codes that are injected in each level and modify second order statistics of the channels, in a spatially invariant manner~\cite{huang2017arbitrary}. The style codes are learned from the outputs of the mapping network, using affine transformation.

\smallskip
We employ a discriminator~\cite{goodfellow2014generative}, $D$, that holds the reverse architecture of the synthesis network. 
It receives generated or real motion, and processes it in skeleton-aware neural blocks that downscale gradually.
A recap of StyleGAN, 
and 
details 
on training setups and hyperparameters,
are given in \ifappendix{\cref{sec:arch_detail,sec:hyperparameters}, respectively.}\else{the sup. mat. }\fi

We train our network with all the StyleGAN losses~\cite{karras2020analyzing}: adversarial loss, path length regularization and $R1$ regularization. For completeness, these losses are detailed in \ifappendix{\cref{sec:gan_losses}. }\else{the sup. mat. }\fi We add two skeleton related losses.

Accurate foot contact is a major factor of motion quality. There is already special care for foot contact in the adversarial loss, as $\mm_{nat}$ contains foot contact labels. However, we noticed that encouraging the contact between the feet and the ground improves the naturalness of the motions, and discourages the phenomenon of ``floating'' feet. Hence, we add an encouragement regulation

\vspace{-5pt}
\begin{equation}
\Loss_{tch}^G  =  
\underset{z\sim \zz}{\E} \left[ - \log s(G(z)_F)) \right], 
\end{equation}
where $(\cdot)_F$ is the contact-label component of the motion, and $s(\cdot)$ is the sigmoid function.

In addition we use contact consistency loss~\cite{li2022ganimator,shi2020motionet}, which requires that a high velocity should not be possible while a foot is touching the ground:

\vspace{-10pt}
\begin{equation}
\Loss_{fcon}^G  =  
\underset{z\sim \zz}{\E} \left[ \norm{FK\left(G(z)\right)_f}_2^2
\cdot s\left(G(z)_F\right)\right], 
\end{equation}
where $FK$ is a forward kinematic operator 
yielding joint locations, and $(\cdot)_f$ is feet velocity extracted from them.

Although our foot contact losses notably mitigate sliding artifacts, we further clean foot contact with a fully automatic procedure using standard IK optimization~\cite{li2022ganimator}.

\subsection{Encoder Architecture} \label{sec:enc_arch}
\vspace{-5pt}
\begin{figure*}
\centering
\includegraphics[width=.9\textwidth]{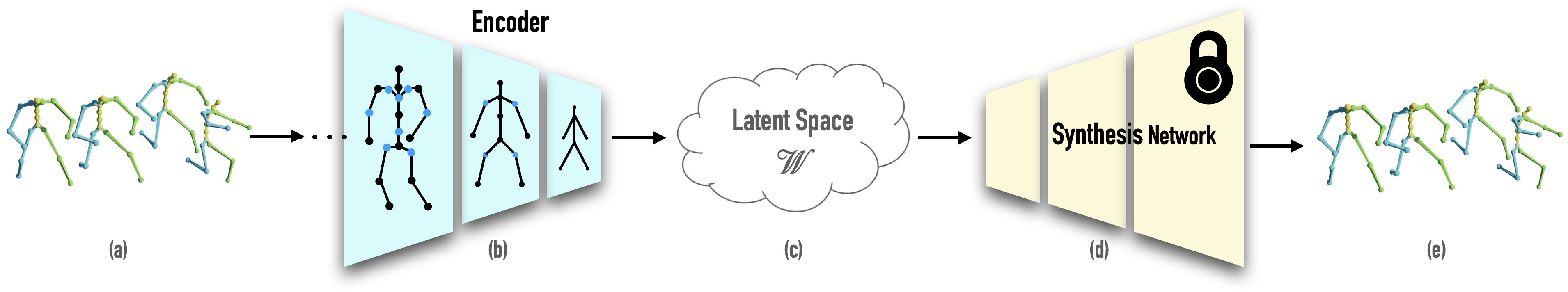} 

\setlength{\abovecaptionskip}{0pt plus 3pt minus 2pt}
\setlength{\belowcaptionskip}{-10pt plus 3pt minus 2pt}

\caption{Our encoder receives an input motion (a), learns a hierarchy of skeleton aware layers (b) and outputs the latent value of that motion (c). 
Once the projected latent value is fed into the synthesis network (d) (with fixed weights), the input motion is reconstructed (e). }
\label{fig:arch_encoder}
\end{figure*}

\begin{figure}
\centering
\includegraphics[width=\linewidth]{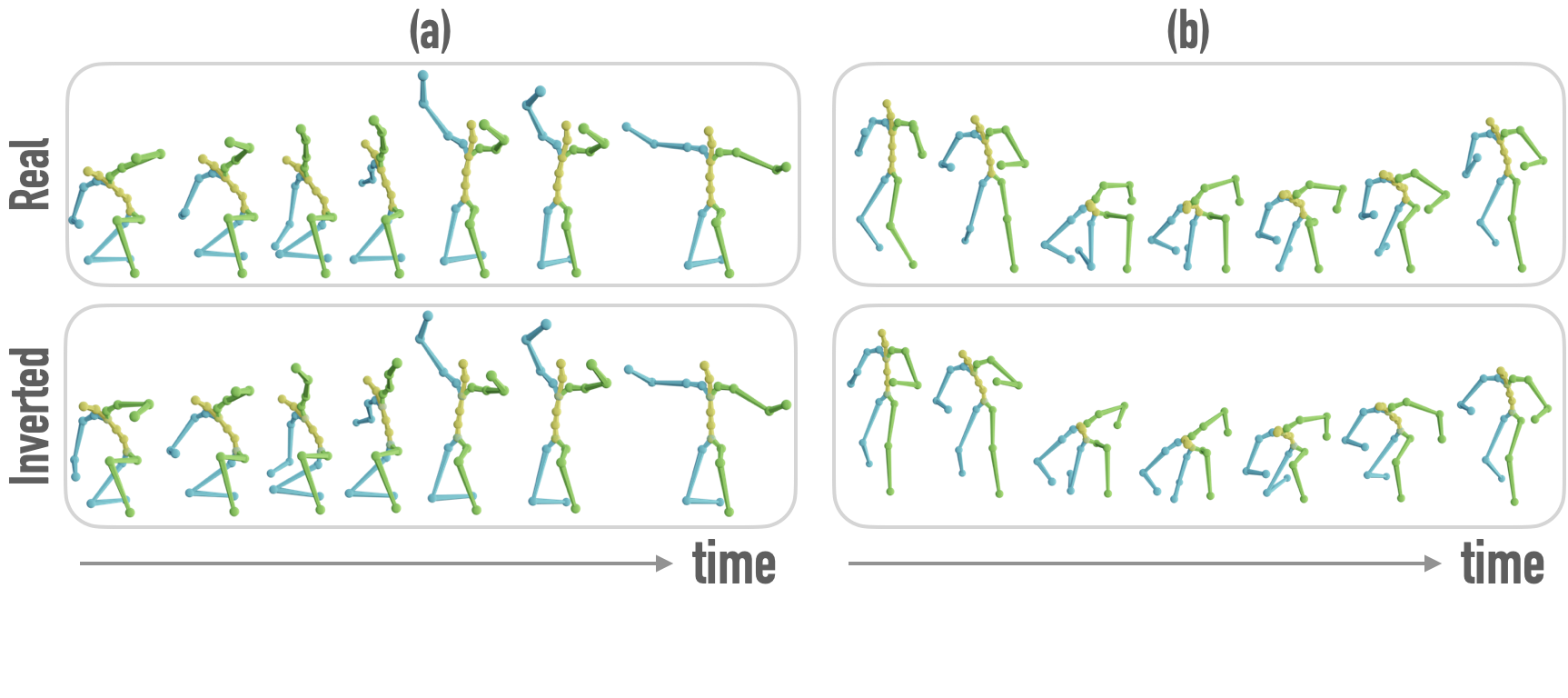} 

\setlength{\abovecaptionskip}{0pt plus 3pt minus 2pt}
\setlength{\belowcaptionskip}{-10pt plus 3pt minus 2pt}

\caption{Inversion of two motions, (a) and (b). 
The original and the reconstructed motions are depicted at the top and bottom rows, respectively.
The reconstruction is done by first projecting the real motions into $\ww+$ using the encoder, and then running the obtained latent values in the generator.
}
\label{fig:encoder}

\end{figure}

Our encoder accepts an input motion and projects it onto the learned latent space such that it can be reconstructed by our synthesis network (with fixed weights), as depicted in  \cref{fig:encoder}.
Note that the input motion can originate from a video footage (using 3D motion reconstruction~\cite{gordon2022flex}), a dataset, or a motion-capture system.
Our encoder enables using the advantages of our well-behaved learned latent space on real motions rather than on generated ones. Moreover, the encoder aims its projections to the ``healthy" part of the latent space, resulting in natural looking results to ill-posed tasks (see \cref{sec:encoder_apps}).

Our encoder, $I$, holds the reverse architecture of the synthesis network, similarly to the discriminator $D$. 
It processes the input motion in skeleton-aware neural blocks that downscale gradually, as shown in \cref{fig:arch_encoder}.
Inversion quality is improved when using $\ww+$~\cite{abdal2019image2stylegan} rather than $\ww$. 
See \ifappendix{\cref{sec:w_plus}. }\else{the sup. mat. }\fi for a recap regarding $\ww+$.

In order to train the encoder we split $\mm_{nat}$ into sets of train $\mm_{trn}$ and test $\mm_{tst}$, with a ratio of 80:20, respectively, and train the encoder on the training set only. The encoder is trained with several losses.
\paragraphtinyvert{Reconstruction loss} The main goal of the encoder is to predict a latent variable $I(m)\in\ww+$ such that $G(I(m))$ is as close as possible to the input motion $m$:
\vspace{-5pt}
\begin{equation}
\Loss_{rec}^I  =  
\underset{m\sim \mm_{trn}}{\E} \left[ \norm{m-G(I(m))}_2^2\right].
\end{equation}
\vspace{-5pt}

\sr{

\paragraphtinyvert{Foot contact loss} 
Unlike the unsupervised foot contact loss of the generator, the following loss~\cite{shi2020motionet} is supervised: 
\begin{equation}
\Loss_{fcon}^I  =  
\underset{m\sim \mm_{trn}}{\E} \left[
BCE(m_F,s(G(I(m)_F))
\right],
\end{equation}
where BCE is a binary cross entropy function, $(\cdot)_F$ is the contact-label component of the motion, and $s(\cdot)$ is the sigmoid function.
}

\paragraphtinyvert{Root loss} 
We noticed that the positions and rotations of the root converge slower than the rotations of the other joints, hence, we add a dedicated loss term:
\vspace{-5pt}
\begin{equation}
\Loss_{root}^I  =  
\underset{m\sim \mm_{trn}}{\E} \left[ \norm{m_{root}-G(I(m))_{root}}_2^2\right], 
\end{equation}
where $(\cdot)_{root}$ are the root velocity and rotation in a motion. This loss is equivalent to putting more weight on the root components in $\Loss_{rec}$.

\paragraphtinyvert{Position loss} 
In addition to the reconstruction loss that mainly supervises rotation angles, we regularize our encoder by a supervision on the joint position themselves:

\vspace{-10pt}
\begin{equation}
\Loss_{pos}^I  =  
\underset{m\sim \mm_{trn}}{\E} \left[ \norm{FK(m)-FK(G(I(m)))}_2^2\right],
\end{equation}
Finally, the total loss applied to the encoder is:

{
\ifAAAI \smallmath \fi
\vspace{-5pt}
\begin{equation}
\Loss^I  =  \Loss_{rec}^I + 
 \lambda_{fcon}^I\Loss_{fcon}^I +\lambda_{root}\Loss_{root}^I + \lambda_{pos}\Loss_{pos}^I, 
\end{equation}
\vspace{-10pt}
}

where we mostly use $\lambda_{fcon} \narroweq 100$, $\lambda_{root} \narroweq 2$,  $\lambda_{pos}\narroweq 0.1$.

\section{Applications}\label{sec:applications}

\subsection{Latent Space Analysis and Applications}\label{sec:latent_space}

\paragraphtinyvert{Latent Clusters} 

\begin{figure}
\centering
\includegraphics[width=\columnwidth]{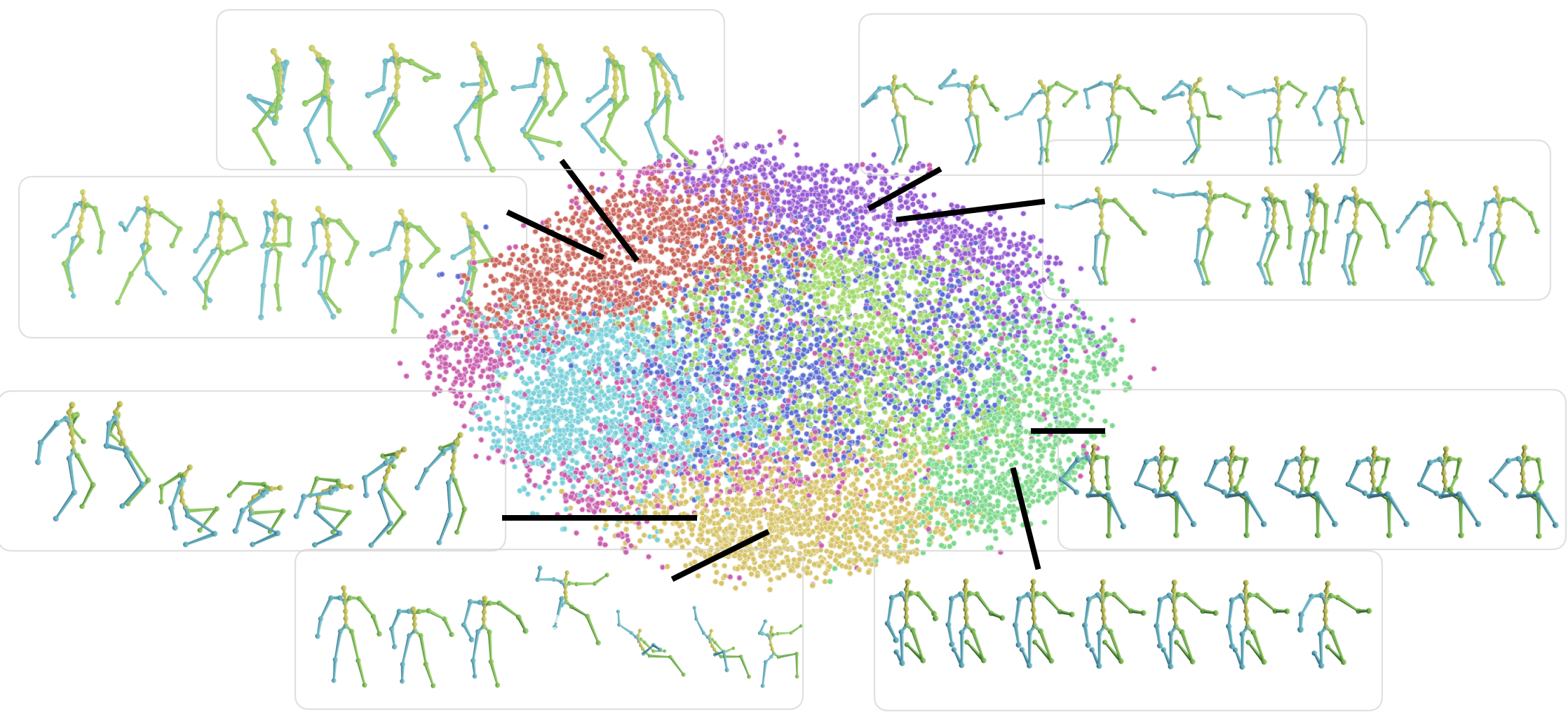} 

\setlength{\abovecaptionskip}{0pt plus 3pt minus 2pt}
\setlength{\belowcaptionskip}{-10pt plus 3pt minus 2pt}

\caption{The latent space $\ww$, split into 8 clusters using K-means, and visualized using T-SNE.
Each point relates to one $\ww$ space instance, generated from random noise $z\in \zz$.
The visualized motions are the result of running these latent variables through our generator $G$.
We observe that the clusters indeed represent semantic grouping of the data.}

\label{fig:k_means_latent}
\end{figure}

We demonstrate that $\ww$ is well-structured by clustering it into meaningful collections of motions. Recall that \algoname{} learns form datasets that  cannot be semantically clustered due to their unstructured nature.

In \cref{fig:k_means_latent} we observe the latent space $\ww$, split into 8 clusters using K-means. The $\ww$ values belong to 10,000 randomly synthesized motions. We randomly choose several motions from each cluster and depict them.
Clearly, motions represented by different clusters are semantically different, and motions that share a cluster are semantically similar.

\paragraphtinyvert{Latent interpolation}

\begin{figure}
\centering
\includegraphics[width=\linewidth]{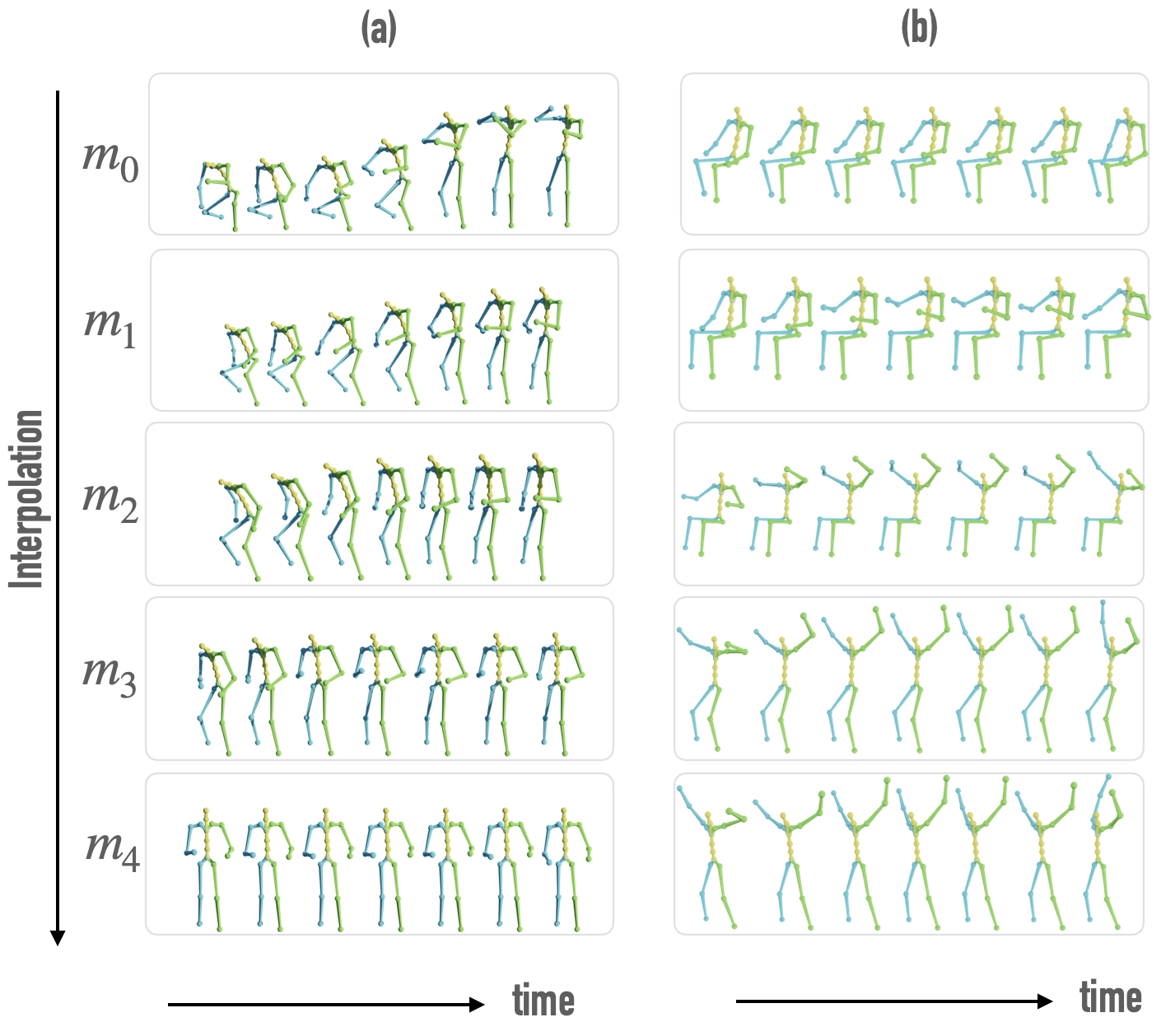} 

\setlength{\abovecaptionskip}{0pt plus 3pt minus 2pt}
\setlength{\belowcaptionskip}{-10pt plus 3pt minus 2pt}

\caption{
Interpolation in the latent space: 
$(a)$ interpolation to the mean motion (truncation); $(b)$ interpolation between sampled motions. Note that the interpolated motions seem natural, even if the interpolation is performed between sitting and standing, a results that can be achieved if the interpolation is performed between the joint positions.
}
\label{fig:interp}

\end{figure}

We demonstrate the linearity of the latent space $\ww$ by
interpolating between the latent values 
and observing the motions generated out of the interpolated values shown in \cref{fig:interp}. A formal definition, detailed descriptions, and analysis of our results are given in \ifappendix{\cref{sec:latent_interp}. }\else{the sup. mat. }\fi

\paragraphtinyvert{Editing in the latent space}

\begin{figure}
\centering
\includegraphics[width=\linewidth]{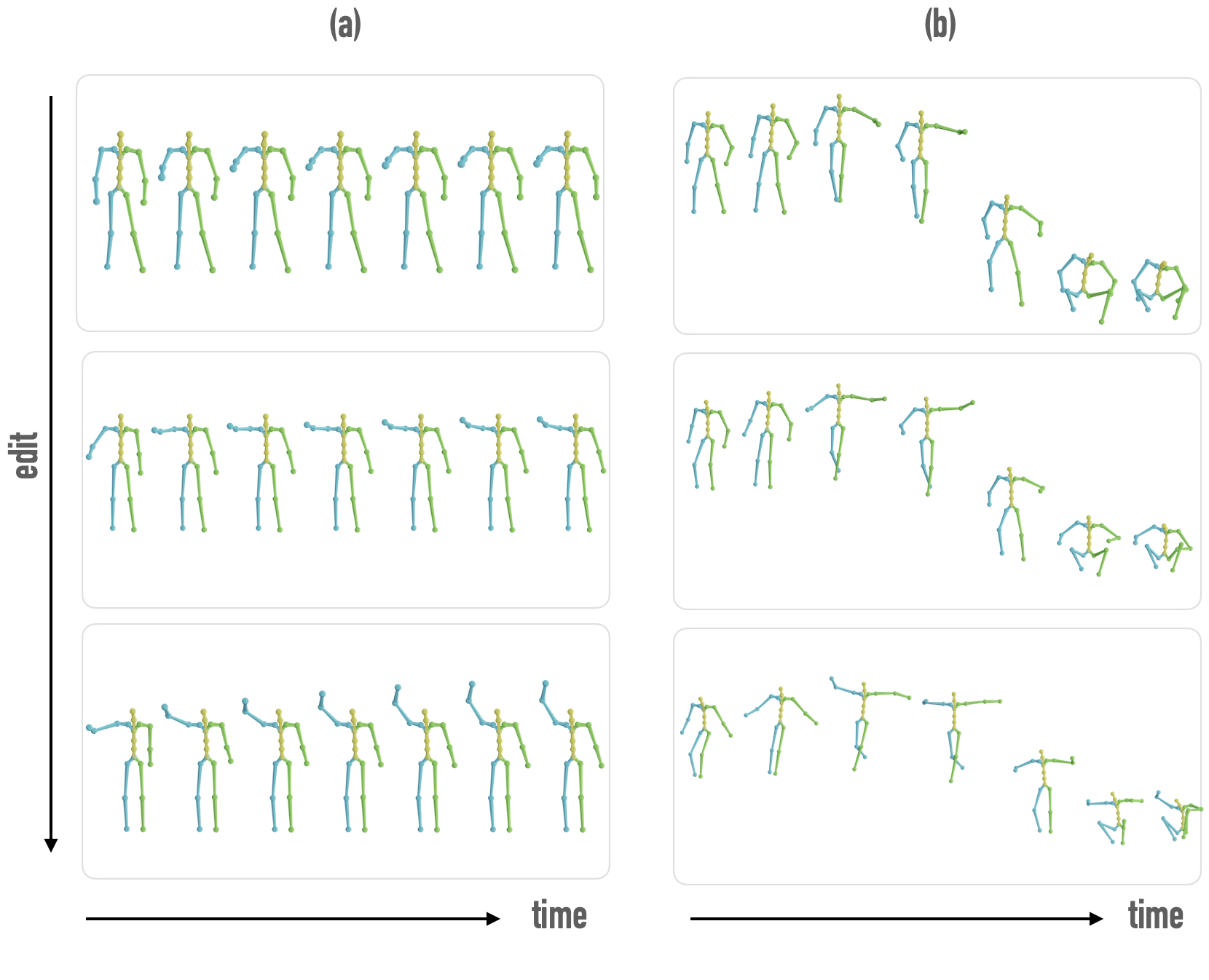} 

\setlength{\abovecaptionskip}{0pt plus 3pt minus 2pt}
\setlength{\belowcaptionskip}{0pt plus 3pt minus 2pt}

\caption{Editing in the latent space. The motion remains intact except for the edited attribute, \emph{gradual right arm lifting} (\emph{gral}).
The \emph{gral}
attribute gets stronger as we go down in each column. 
The generative prior of MoDi keeps the jumping motion (b) natural, even at the expense of arm lifting.
}
\label{fig:edit}
\end{figure}

If a latent space is sufficiently disentangled, it should be possible to find direction vectors that consistently correspond to individual factors of variation. Let $a$ be an attribute related to motion. $a$ can be any semantic attribute, such as motion speed, verticality measurement of parts in the body, or a motion style. 
Inspired by Shen \etal \shortcite{shen2020interfacegan}, we compute a score that measures 
$a$ 
in a motion. 
For example, when measuring the verticality of a motion, a character doing a handstand would get a score of $-1$, lying down would get a score of $0$, and standing up would get a score of $1$. 
Using a given score, we train an SVM, yielding a hyperplane that serves as a separation boundary. 
Denote the unit normal of the hyperplane by $n$. Then $G(w+n)$ possesses increased score of attribute $a$ comparing to $G(w)$.
The only attribute that should change in such editing is $a$, preserving the rest of the motion intact.

Unlike image datasets, that hold labeling for various attributes (age, gender,...), there is not much labeling in motion datasets.
We create our own simple classifiers, and elaborate next regarding one of them, measuring \emph{gradual right arm lifting}, denoted \emph{gral}. \emph{Gral} means that the right arm is lifted as time advances.
Computing a score for the \emph{gral} attribute is not straightforward, and is detailed in\ifappendix{ \cref{sec:gral_score}. }\else{the sup. mat. }\fi 
Our results are visualized in \cref{fig:edit}, where we show that the \emph{gral} attribute gets stronger while stepping in the latent space, and the naturalness of motions as well as their semantics are kept. 
In our video clip we show that when such an attribute is artificially applied via geometric interpolation, the results are unnatural. Obtaining manual natural results would require an artist's hard work.

\paragraphtinyvert{Crowd simulation}

\begin{figure}
\centering
\includegraphics[width=\linewidth]{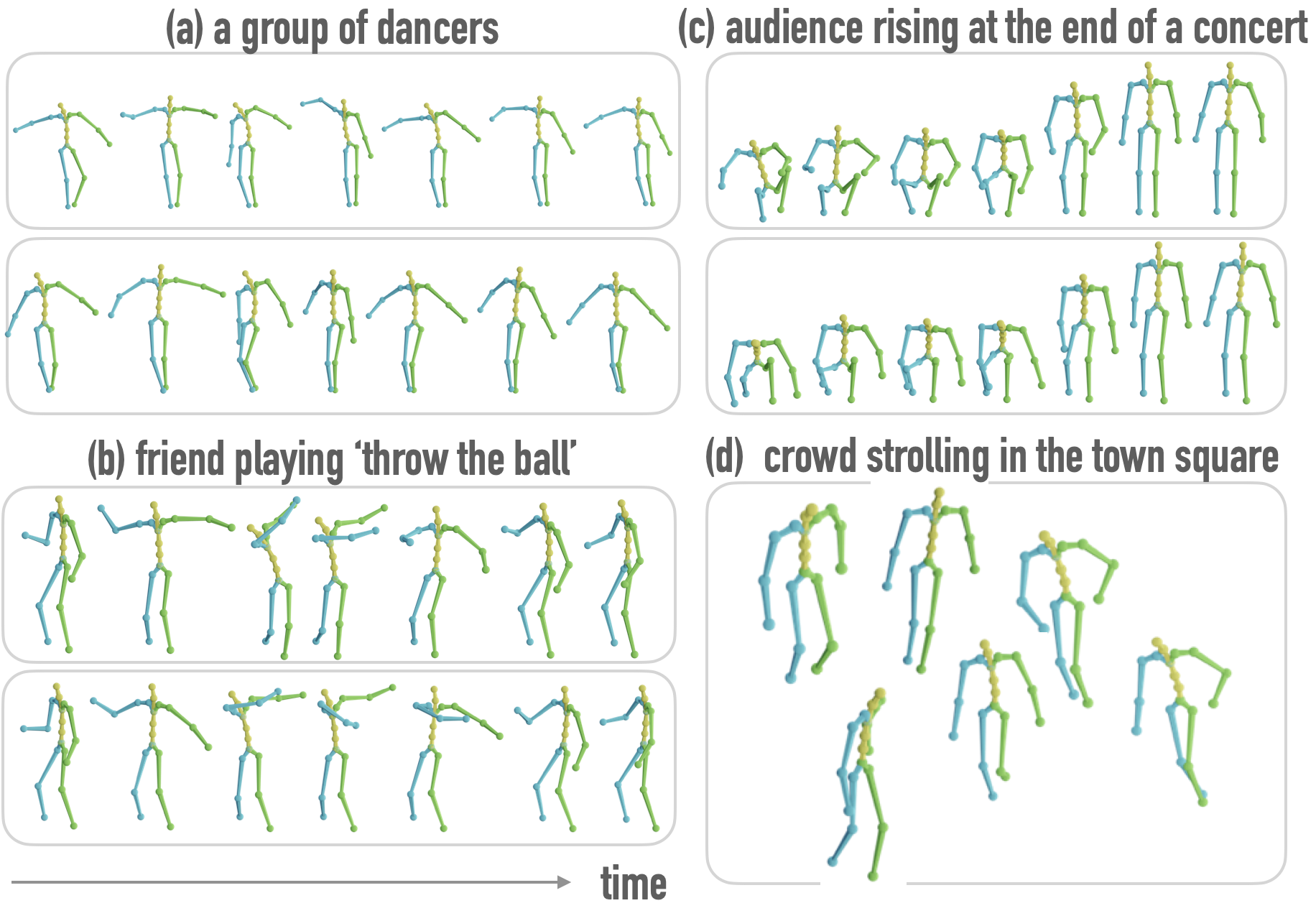} 

\setlength{\abovecaptionskip}{0pt plus 3pt minus 2pt}
\setlength{\belowcaptionskip}{-5pt plus 3pt minus 2pt}

\caption{Crowd simulation. Blocks (a), (b) and (c) depict sequences of motion frames over time. The two sequences in each of these blocks visualise similar motions created using perturbation in the latent space. Block (d) depicts poses in \emph{one} time frame extracted from six \emph{distinct} motions.
}
\label{fig:crowd}
\end{figure}
Given an input motion $m$, we can sample variations in our latent space $\ww$ by simply sampling the neighborhood of the $w$ that corresponds to $m$. 
We sample in a Gaussian $ \! \sim \!\nn(m,\sigma^2)$, with $\sigma$ in the range 0.1-0.8. This way, one can simulate, for instance, people walking in the town square, a group of dancers, or a group of friends jumping from a bench. See \cref{fig:crowd} and our video for examples.

\subsection{Solving Ill-posed Tasks with the Encoder} \label{sec:encoder_apps}
\vspace{-5pt}

\begin{figure}
\centering
\includegraphics[width=\columnwidth]{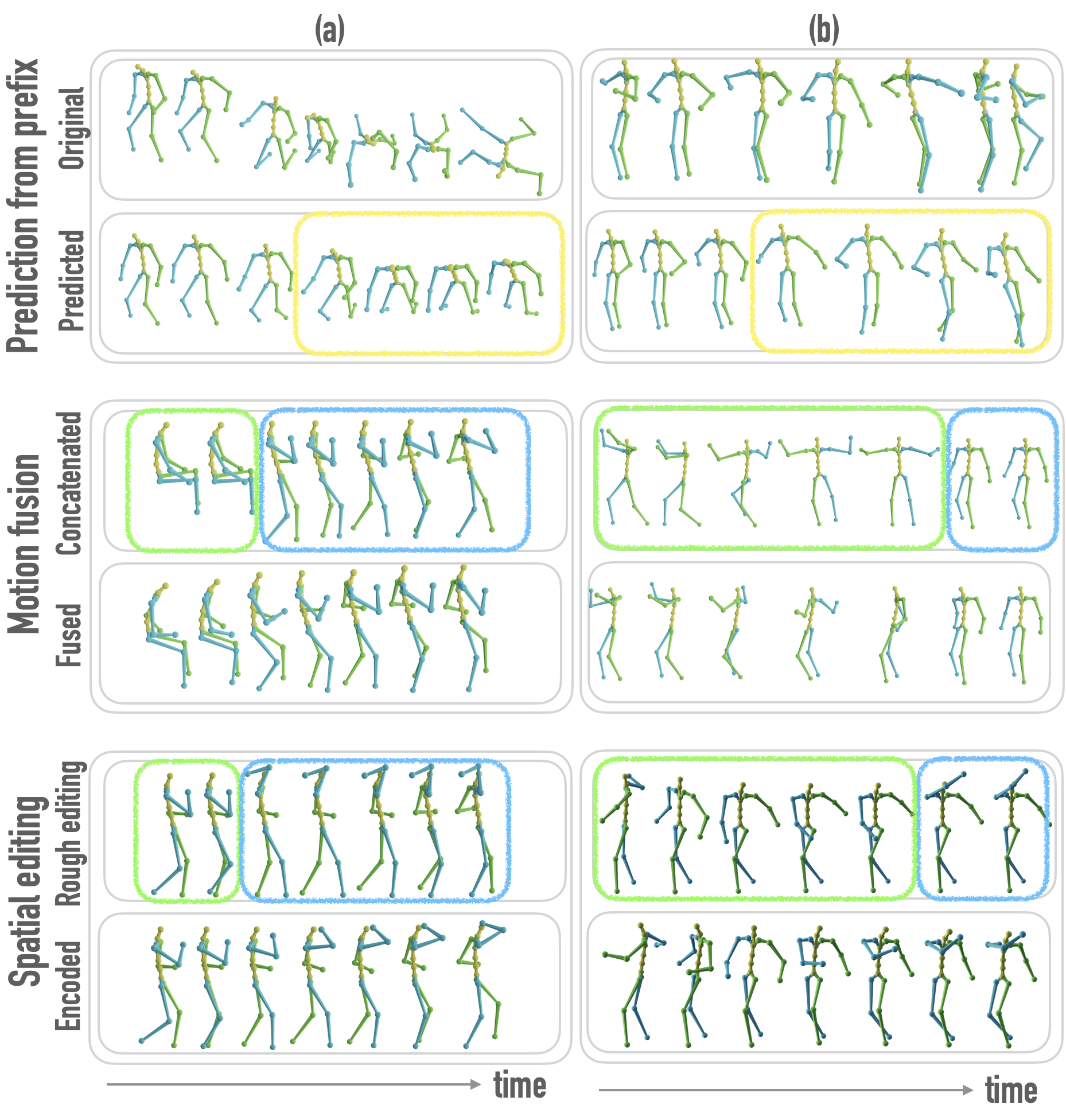} 

\setlength{\abovecaptionskip}{3pt plus 3pt minus 2pt}
\setlength{\belowcaptionskip}{-10pt plus 3pt minus 2pt}

\caption{
\sr{
Our encoder enables coping with ill-posed tasks. 
Prediction from prefix: the top row is the original motion, from which only the prefix is seen by the encoder. The predicted output is in the second row. Notice that the encoder synthesises a coherent suffix (in yellow), without overfitting the original one.
Motion fusion: (a) sitting (green) and walking (blue). Notice the smooth fused sequence versus the concatenated one; (b) dancing (green) and standing (blue) where the last green frame looks backwards, while the first blue frame looks forward. The encoder mitigates this challenging concatenation by gradually rotating the character.    
Spatial editing: green and blue rectangles encircle frames left intact or amateurishly edited, respectively.
Here we edit the \emph{gradual right arm lifting (gral)} attribute using a very different approach. Instead of going through the tedious work of finding an editing direction in the latent space (\cref{sec:latent_space}), we spatially edit the right arms and let the encoder turn our rough edit into a natural one. 
}
}
\label{fig:encoder_apps}

\end{figure}

Many tasks in the motion domain are ill-posed. That is, there is more than one single solution to the task.
We next present several examples, each of which using an encoder yields high quality results to ill-posed tasks. 
The tasks in this section are conditioned, and show that a generator that has been trained in an unconditional setting can be used for a variety of downstream tasks, as illustrated in ~\cref{fig:encoder_apps}.

All but the first of the following applications use the encoder as is, with no further training. The main idea in these applications is to use the encoder to invert a given arbitrary motion $m$, into the healthy regions of the latent space (this happens with no additional effort), and then generate the motion using our generator. The result $G(I(m))$ is faithful to the input motion, but at the same time, it is projected into the manifold of natural motions. The fine output is imputed to the well-behaved nature of the latent space combined with the generator's ability to generate natural motions only.

In the following paragraphs, we denote motion frame index by $t\in[1\dots T]$. Recall $T$ is the number of frames in a full motion.
\sr{
\paragraphtinyvert{Prediction from Prefix}
Given a motion prefix of length $t$ (namely, $t$ frames that represent the beginning of a motion sequence), complete the motion into length $T$. 
Unlike the other tasks, for this task we re-train our encoder such that the input is zeroed in frames $[(t+1)\dots T]$. The encoder is supervised by the full motion, using its full set of losses. 
Generalizing this solution to another ill-posed task, in-betweening, is straightforward.

\paragraphtinyvert{Motion Fusion}
Given two input motions, take [1\dots t] frames from the first and $[(t+1)\dots T]$ frames from the second, and output one motion whose prefix is the first motion and suffix is the second one. A na\"ive concatenation of the frames would lead to a non continuous motion at the concatenation frame. %
Denote the concatenated motion by $m$. Applying $G(I(m))$ smooths the frames around the concatenation area and yields a natural looking motion, while fidelity to the prefix and suffix motions is kept. 
Generalization to a variable number of motions is simple.

\paragraphtinyvert{Denoising}
Given a noisy input motion $m$, output a natural looking, smooth motion. Similar to motion fusion, we apply $G(I(m))$ and obtain a denoised output.
Generalization to more types of corrupted motions is straightforward.

\paragraphtinyvert{Spatial Editing}
Given an input motion $m$, with several frames manually edited, output a natural looking, coherent motion. 
In the animation industry, often an animator is interested in a spatial change of an existing motion, such that the character performs an action in some frames, \eg, raise a hand. Manually editing these frames in the spatial domain is exhaustive and requires a professional. We tackle this task by performing a bulky manual spatial edit, yielding a non-natural motion, and running it through $G\circ I$ to get a natural and coherent output.
}
\section{Experiments} \label{sec:experiments}
\vspace{-5pt}

\paragraphtinyvert{Datasets} 
We use Mixamo~\cite{mixamo} 
and HumanAct12~\cite{guo2020action2motion},
as elaborated in \ifappendix{\cref{sec:datasets}.}\else{our sup. mat. }\fi

\subsection{Quantitative Results} \label{sec:quantitative}
\vspace{-5pt}
\paragraphtinyvert{Metrics}
We use the metrics FID, KID,  precision-recall and diversity, and describe them in \ifappendix{\cref{sec:metrics}. }\else{the sup. mat. }\fi 
The metrics build upon the latent features of an action recognition model.
However, training such a model on Mixamo is challenging, as there is no action labeling in it.  
We resort to a creative solution, as detailed in \ifappendix{\cref{sec:metrics}. }\else{the sup. mat. }\fi

\paragraphtinyvert{Results}
\begin{table}[t!] 
\centering
\resizebox{\columnwidth}{!}{
\begin{tabular}{*{5}{c}}
\toprule
\textbf{Model}  & \textbf{FID $\downarrow$} & \textbf{KID $\downarrow$} & \textbf{\makecell{Precision $\uparrow$ \\ Recall $\uparrow$}} & \textbf{Diversity $\uparrow$ }\\
\midrule
ACTOR~\shortcite{petrovich2021action}  & 48.8 & 0.53 & \textbf{0.72},   0.74 & 14.1 \\
\midrule
MDM~\shortcite{tevet2022human}  & 31.92 & 0.96 & 0.66, 0.62 & 17.00 \\
\midrule
{\color{gray}\makecell{\algoname{} (ours) \\ with mixing}} & {\color{gray}15.55} & {\color{gray}0.14} & {\color{gray}\textbf{0.72}, 0.75} & {\color{gray}17.36} \\
\midrule
\makecell{\algoname{} (ours) \\ without mixing} & \textbf{13.03} & \textbf{0.12} & 0.71,   \textbf{0.81} & \textbf{17.57} \\
\bottomrule
\end{tabular}

} %

\setlength{\abovecaptionskip}{5pt plus 3pt minus 2pt}

\caption{Quantitative results for state-of-the-art works on the HumanAct12
dataset. 
The grayed line shows our original algorithm, without the changes that make it comparable. Note that our model leads in all the variations. Best scores are emphasised in \bf{bold}.
}
\label{tab:comparison}
\end{table}

\sr{We compare our model with state-of-the-art synthesis networks, ACTOR~\cite{petrovich2021action} and MDM~\cite{tevet2022human}, on the HumanAct12~\cite{guo2020action2motion} dataset, and show the results in \cref{tab:comparison}. 
Note that the compared works use state-of-the-art recent techniques, namely transformers and diffusion models: ACTOR uses transformers and MDM uses diffusion models. 
Yet, in the challenging setting of unconstrained synthesis, \algoname{} outperforms them by a large margin.
To enable a comparison in an unconstrained setting, we re-train ACTOR by assigning the same 
label to all motion instances, and use MDM's unconstrained variation. 
For the sole purpose of comparison with other works, we provide a  version of \algoname{}, that skips style-mixing~\cite{karras2020analyzing}, as
mixing may change the distribution of synthesised motions, yielding degradation in the metric score. Both versions are shown in \cref{tab:comparison}.
}

\subsection{Qualitative Results} \label{sec:qualitative}
\vspace{-5pt}
\begin{figure}[t!]
\centering
\includegraphics[width=\linewidth]{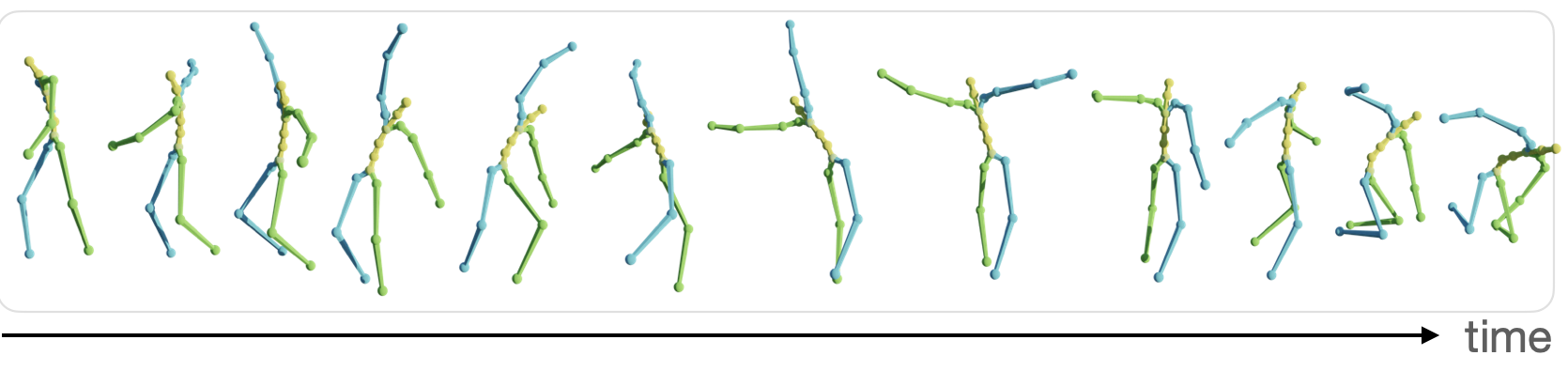} 

\setlength{\abovecaptionskip}{0pt plus 3pt minus 2pt}
\setlength{\belowcaptionskip}{5pt plus 3pt minus 2pt}

\caption{Qualitative result. More are in the video and \ifappendix{in \Cref{sec:qual_sup}.}\else{sup.}\fi}
\label{fig:qual_main}
\end{figure}

The reader is encouraged to watch our supplementary video in order to get the full impression of the quality of our results. For completeness, we show one special motion in \cref{fig:qual_main}, and several more in \ifappendix{\cref{sec:qual_sup}.}\else{the sup. mat. }\fi

\begin{table}[t] 
\centering
\resizebox{\columnwidth}{!}{
\begin{tabular}{*{5}{c}}
\toprule
\textbf{\makecell{Architecture \\ variation}}  & \textbf{FID $\downarrow$} & \textbf{KID $\downarrow$} & \textbf{\makecell{Precision $\uparrow$ \\ Recall $\uparrow$}} & \textbf{Diversity $\uparrow$ }\\
\midrule
\makecell{non skel.-aware} & $23.0^{\pm 0.3}$ & $0.17^{\pm 0.02}$ & \makecell{$0.46^{\pm 0.01}$ \\ $0.41^{\pm 0.01}$} & $13^{\pm 0.08}$ \\
\midrule
\makecell{joint loc.  \\ rather than rot.} & $17.3^{\pm 0.06}$ & $0.2^{\pm 0.03}$ & \makecell{$0.46^{\pm 0.02}$ \\ $0.58^{\pm 0.01}$} & $14.0^{\pm 0.3}$ \\
\midrule
\makecell{pool rather than \\ conv. scaler}  & $14.9^{\pm 0.7}$ & $0.16^{\pm 0.02}$ & \makecell{\pmb{$0.49^{\pm 0.01}$} \\ $0.58^{\pm 0.03}$}  & $15.3^{\pm 0.02}$ \\
\midrule
\makecell{remove one \\ in-place  conv. \\ per hierarchy} & $14.1^{\pm 1.4}$ & $0.15^{\pm 0.02}$ & \makecell{$0.46^{\pm 0.02}$ \\ $0.66^{\pm 0.1}$} & \pmb{$15.4^{\pm 0.9}$}\\
\midrule
\makecell{final architecture} & \pmb{$11.5^{\pm 0.9}$} & \pmb{$0.1^{\pm 0.01}$} & \makecell{$0.46^{\pm 0.02}$ \\   \pmb{$0.69^{\pm 0.02}$}} & \pmb{$15.4^{\pm 0.2}$} \\
\bottomrule
\end{tabular}

} %

\setlength{\abovecaptionskip}{5pt plus 3pt minus 2pt}
\setlength{\belowcaptionskip}{-10pt plus 3pt minus 2pt}

\caption{Quantitative results for various generator designs, on the Mixamo dataset. 
Best scores are emphasised in \textbf{bold}.
}
\label{tab:ablation_gen_design}
\end{table}

\subsection{Ablation}
\vspace{-5pt}

In \cref{tab:ablation_gen_design} we show the results of a thorough study of potential architectures.
We first show the metric scores when using a non skeleton-aware architecture. That is, when representing motion by pseudo images. The drawbacks of pseudo-images are detailed in \cref{sec:related_work}.
In the second study, we use joint locations rather than rotations. In \ifappendix{\cref{sec:motion_rep}} \else{the sup. mat. }\fi we describe why generating rotations is better than generating locations.
Our third study refrains from using our new \filtername{}, and uses skeleton-aware pooling~\cite{aberman2020skeleton}, testifying that our new filter improves the results.
Next, we check \ilreplace{what happens when}{the effect of} removing one in-place convolution from each hierarchical layer. 
Finally, we measure the scores for our final architecture, and conclude that our architectural choices outperform other alternatives.

Every training configuration is run 3 times, and every evaluation (per configuration) 5 times. 
The numbers in \cref{tab:ablation_gen_design} are of the form $mean^{\pm std}$.
More ablation studies can be found in \ifappendix{\cref{sec:more_ablation}. }\else{the sup. mat. }\fi

\vspace{-5pt}
\section{Conclusion}
\vspace{-5pt}

One of the most fascinating phenomena of deep learning is that it can gain knowledge, and even learn semantics, from unsupervised data. In this work, we have presented a deep neural architecture that learns motion prior in a completely unsupervised setting.
The main challenge has been to learn a generic prior from a diverse, unstructured and unlabeled
motion dataset. This necessarily requires a careful design of a neural architecture to process the unlabeled data. We have presented \algoname{}, an architecture that distills a powerful, well-behaved latent space, which then facilitates downstream latent-based motion manipulations.

Like any data-driven method, the quality of the generalization power of \algoname{} is a direct function of the training data, which, at least compared to image datasets, is still lacking. Another limitation is that skeleton-aware kernels, with dedicated kernels per joint, occupy large volumes, resulting in relatively large running time.  

In the future, we would like to address the challenging problem of learning motion priors from video.
Building upon networks like \algoname{}, with proper inductive bias, may open the way towards it.

\ifanonymous
\else
    \section{Acknowledgments}
This work was supported in part by the Israel Science Foundation (grants no. 2492/20 and 3441/21).

\fi

\ifappendix
    \newpage
    \centerline{\LARGE\bfseries Appendix\par}
    \appendix
    \section{Outline}

This \ifappendix{Appendix }\else{Supplementary }\fi  adds details on top of the ones given in the main paper. While the main paper stands on its own, the details given here may shed more light. The majority of this Supplementary recaps existing algorithmic elements that are used by our work.

In \cref{sec:model_detail} we provide more details regarding our model; considerations that led us to choosing 3D convolutions and edge rotation representation, description of skeleton aware models in existing works, and more details on the StyleGAN architecture.
\cref{sec:app_details} describes the $\ww+$ space used for inversion, and the computation of the \emph{gral} score, used for latent motion editing.  
Lastly, in \cref{sec:experiments_detail} we elaborate on our experiments; we describe the datasets that we use, provide implementation details such as hyper-parameters, detail metrics, and show additional ablation and additional qualitative results.

\section{Model -- Additional Details} \label{sec:model_detail}

\subsection{Structure-aware Neural Modules -- Additional Details} \label{sec:neural_modules}

In this section we first describe our motivation in using 3D convolutions;
then, we depict the way these convolutions are used by our networks. 
Finally, for information completeness, we describe skeleton aware neural modules from existing works, and the way we use some of them in a 3D convolutional setting.

\paragraph{3D convolutions -- motivation} 
Described below are two ways to design the filters of \emph{skeleton-aware} convolutions~\cite{aberman2020skeleton}.
Recall that $E$, $T$, $\ell$, and $U$ denote the number of entities and frames, the hierarchical level index, and the kernel width, respectively.
\begin{itemize}
    \item 
    Existing works use 3D filters of dimension $(K_{\ell+1} \cdot E_{\ell+1}) \narrowtimes (K_\ell  \cdot E_\ell) \narrowtimes U $. 
    Such filters are applied by the neural model using 1D or 2D convolutions, over the time axis or time and joint-channels axes, respectively.
    \item 
    Our work uses 5D filters of dimension $K_{\ell+1} \narrowtimes K_\ell \narrowtimes E_{\ell+1} \narrowtimes E_\ell \narrowtimes U $. 
    These filters are applied by the neural model using 3D convolutions, over the time and joint axes, and an additional axis that holds the output joints, to be described next.
\end{itemize}

The convolutions in existing works combine the joints and the channels into the same dimension, yielding a non intuitive representation that adds complexity to coding. For example, the output joints are received as channels, and require reshaping to be represented as joints. It is common practice in most neural architectures to hold a dedicated dimension for the channels.
Moreover, 3D filters introduce complication when combined with the StyleGAN algorithm, for two distinct reasons:
\begin{enumerate}
    \item 
    StyleGAN uses modulation, which is difficult to apply if the channels and the joints share the same dimension, as the style is injected to each channel separately (see \cref{sec:arch_detail}). By using 3D convolutions, \ie 5D filters, we place the channels in their own dedicated dimension, so modulation becomes simple.
    \item
    StyleGAN uses transposed convolutions, in which the axes are swapped such that the output and input channels switch places. Managing such a swap becomes straightforward when the channel dimensions are separated from the joint dimensions.
\end{enumerate}

Note that it \emph{is} possible to keep using 3D filters as done in other works. However, such usage, combined with StyleGAN's components, adds complexity (multiple data reshapes, weights reshapes, and dimension swaps) to those implementing the model.

\paragraph{3D convolutions -- usage}
\begin{figure*}
\centering
\includegraphics[width=.9\linewidth]{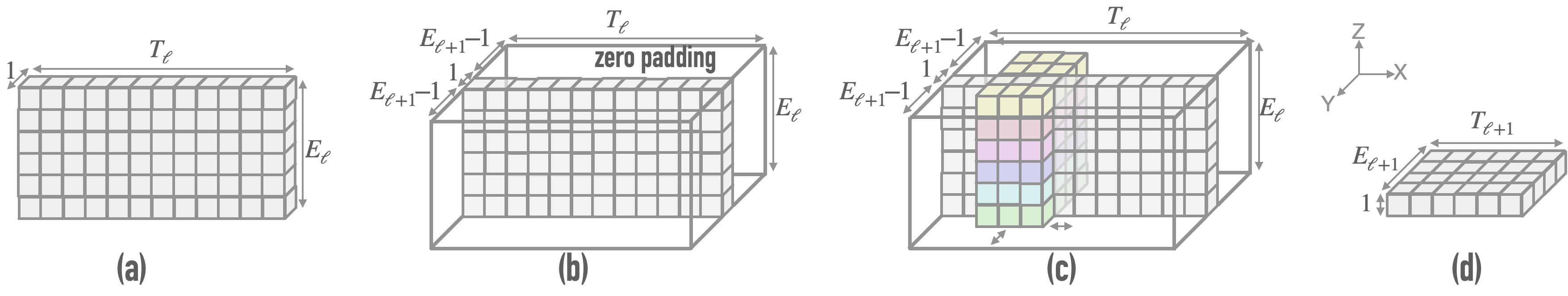}

\caption{
\sr{
Down sampling with our new filter
(depicted 3D out of 4D, no channels):
(a) Data at hierarchical level $\ell$: Dimensions are $K_\ell \times E_\ell \times T_\ell$. We expand it by one dimension in preparations for 3D convolution.
(b) Level $\ell$ data is further padded by zeros, and its new dimensions are $K_\ell \times (2 E_{\ell+1}\!\!-\!\!1) \times E_\ell \times T_\ell$.
(c) 3D convolution: The filter is slid within the data block. Sliding is along the x and y axes only, as the z axis' filter height is identical to the data height.
(d) Resulting data: The extra dimension of size $1$ is dropped such that final dimensions  at level $\ell+1$ are $K_{\ell+1} \times E_{\ell+1} \times T_{\ell+1}$.
}}
\label{fig:conv_weight}
\end{figure*}

\Cref{fig:conv_weight} describes the way in which our networks use 3D convolutions. 
As explained in the main paper,
we dedicate separate kernels to each joint. To convolve each separate kernel with the data, we use different dimensions for the input and output joints.
The output joints axis is created by expanding the data by one dimension followed by zero padding, such that sliding the filters along the new axis enables using different weights for each output joint. Once convolution is completed, the result holds the joints data in the output joint axis, and the input joint axis becomes degenerated (of size 1) and is removed.

\paragraph{Recap existing neural modules}
The modules described here are skeletal in-place convolution and skeletal pooling~\cite{degardin2022generative,yu2020structure,yan2019convolutional,aberman2020skeleton}.  
In our work we create a 3D version of the skeletal in-place convolutional filter, and replace the skeletal pooling by our novel \filtername{} filter.

In \cref{fig:pool} we show a skeletal pooling procedure. Pooling is done by averaging the features of two entities, hence, it is equivalent to a convolution with weights of 0.5. Our new filter applies a convolution with learned weights, generalizing the pooling functionality, and allowing the network the freedom to choose the optimal weights.

In \cref{fig:conv_inplace} we depict our 3D version of a skeleton-aware convolutional filter. Unlike our novel \filtername{} filter, this filter is an in-place one, which means it retains the dimensions of its input, and cannot scale it.

\begin{figure}
\centering
\includegraphics[width=.7\linewidth]{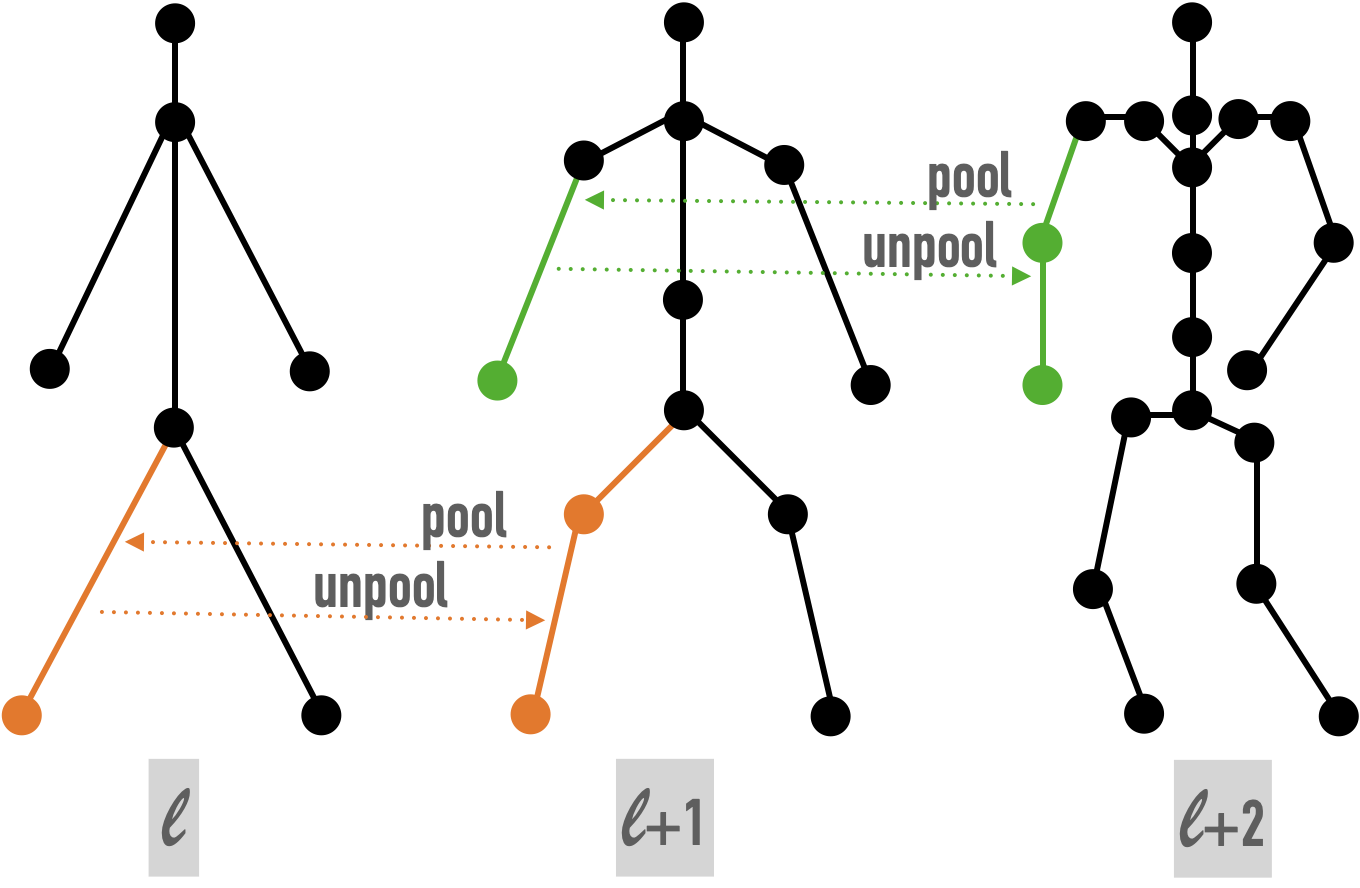}

\caption{
Skeletal pooling, not used by MoDi. The pooling operation merges two adjacent edges and removes the joint between them. The unpooling operation splits an edge into two, and adds a joint between the newly created edges.
We denote skeletal hierarchy levels with $\ell,\ell+1,\ell+2$, and demonstrate pooling and unpooling on selected joints in orange (levels $\ell,\ell+1$), and in green (levels $\ell+1,\ell+2$).}
\label{fig:pool}
\end{figure}

\begin{figure}
\centering
\includegraphics[width=\linewidth]{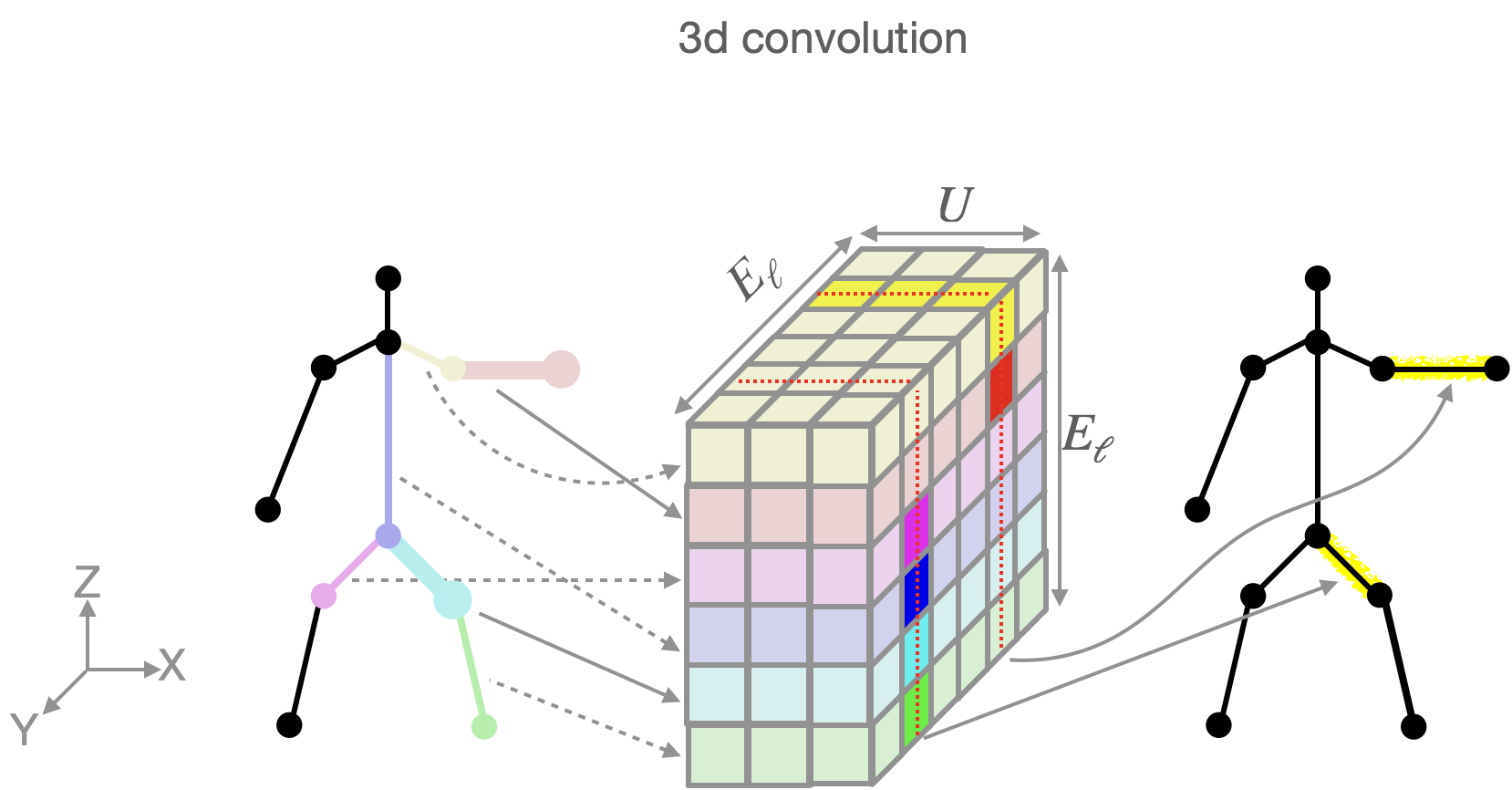}

\caption{
Our 3D version of a skeleton-aware in-place convolutional filter. %
Each horizontal slice (xy plane) is related to one entity in the input character (left), and each vertical slice (xz plane) is related to one entity in the output character (right). 
Each entity in the output character ``sees" only weights related to its neighboring entities, emphasised with saturated colors in the filter.  
We demonstrate convolutions on the left thigh and on the left forearm, marked yellow in the output character. Note that each of these entities is affected by its immediate neighbors and ignores entities that do not neighbor it.
Our filter is 5D and since we can only visualize 3D, we omit the channels.
Recall that $E$, $\ell$, and $U$ denote the number of entities, the hierarchical level index, and the kernel width, respectively.
}
\label{fig:conv_inplace}
\end{figure}

\subsection{Motion Representation Considerations} \label{sec:motion_rep}
Some methods generate a sequence of 3D poses~\cite{degardin2022generative}
, where each
location is specified by the 3D coordinates of each joint. However, the resulting representation is incomplete, since it does not reflect a rotation of a bone around its own axis. In particular, it does not contain all the information necessary to drive a rigged virtual 3D character, and the temporal consistency of the skeleton’s bone lengths is not guaranteed. While joint rotations may be recovered from joint positions via inverse kinematics (IK), the solution is not unique,
and thus ill-posed. 
Furthermore, models that predict positions tend to yield a temporally jittery output, and require a post processing smoothing stage.
Due to these considerations, we follow numerous recent works that are based on joint rotation representation~\cite{maheshwari2022mugl,li2022ganimator}. The motion generated by \algoname{} can be directly converted into an animation sequence without the need to apply neither IK nor temporal smoothing.

Our network is trained on a single set of bone lengths. Once a motion is generated, it can be retargeted to any other set of bone lengths using existing motion retargeting methods~\cite{aberman2020skeleton, aberman2020unpaired, biswas2021hierarchical}.

\subsection{Generative Network Architecture in Detail} \label{sec:arch_detail}

In this section we provide further details regarding the architectural building blocks of \algoname{}. Some of the description is based on StyleGAN~\cite{karras2020analyzing} and is given here for information completeness.

\paragraph{Generator}
\begin{figure*}
\centering
\includegraphics[width=\linewidth]{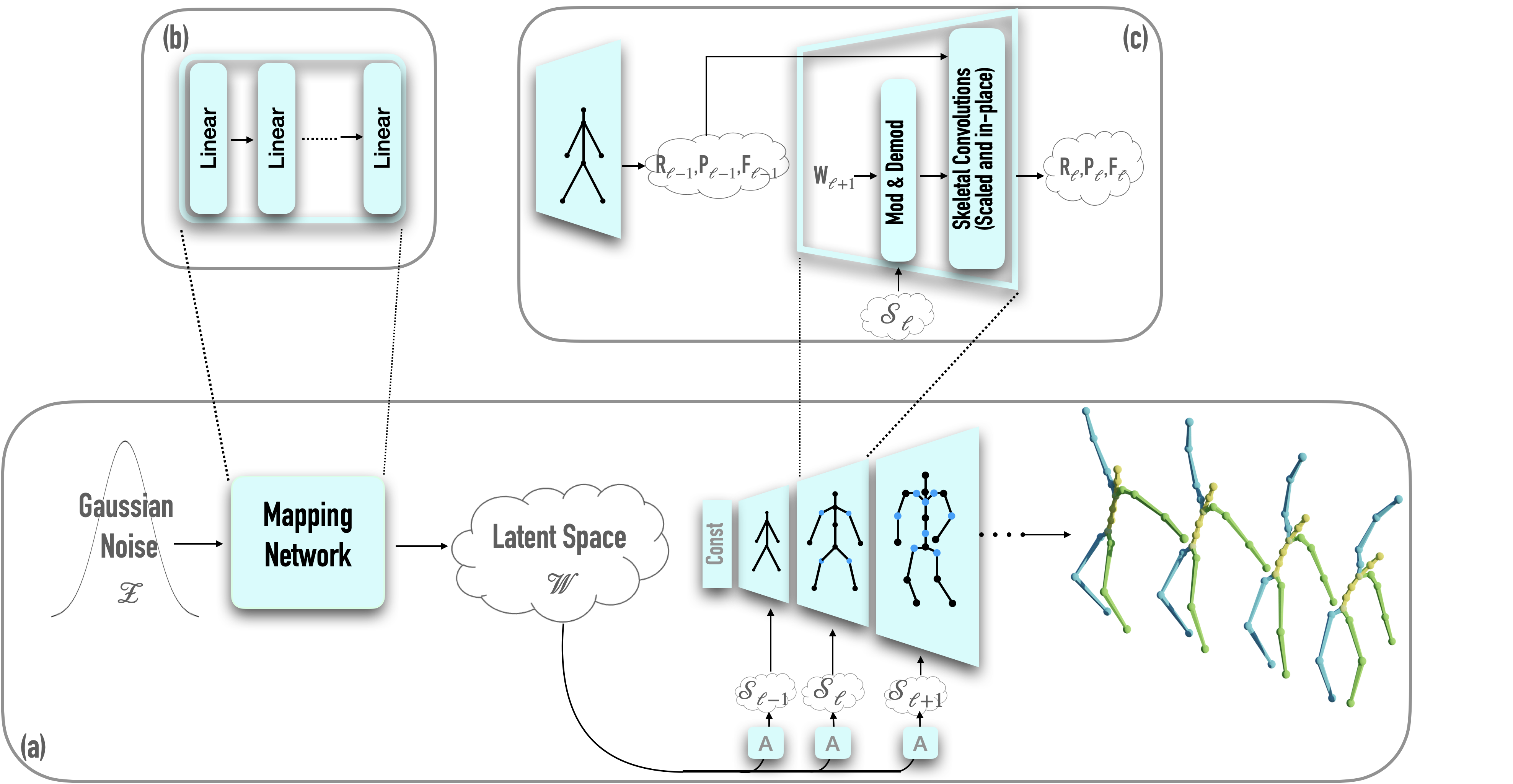} 

\caption{Our motion generator in detail. 
(a) Adding style injection information to the architecture depicted in 
the main paper. $\boxed{A}$ denotes a learned affine transformation. This transformation is applied on the latent code $w$ to produce a style code $\mss_\ell$, where $\ell$ is the hierarchical level index. A different style code is injected to each layer. (b) Zoom in on the mapping network, which is an MLP with several linear layers. (c) Zoom in on the motion synthesis network, where a style code $\mss$ modulates the layer's weight. The styled weight is then used for a transposed convolution of the layer features. Recall that $\bR_\ell$, $\loc_\ell$  and $\bF_\ell$ denote the features in level $\ell$ of rotations, root positions and foot contact labels, respectively. A transposed skeletal convolution applies the modulated weights on the data features from the previous (coarser) hierarchical level. Since the convolution is transposed, it results with larger dimensions, both in the temporal axis and in the joints axis. 
}
\label{fig:arch_detail}
\end{figure*}

In \cref{fig:arch_detail} we show additional details related to the motion generator. In particular, we depict the usage of modulation and demodulation~\cite{karras2020analyzing}, which has been shown to be safer compared to AdaIN~\cite{huang2017arbitrary} in terms of certain artefacts. 
The AdaIN block processes \emph{data}, namely normalizes it and applies a new standard deviation.
The modulation/demodulation block performs an equivalent (in expectation) operation on the \emph{weights}. 
Let $u$ denote a weight value within a filter, and let $i$, $j$ and $k$ denote the input channel index, output channel index, and filter spatial index, respectively.
Instead of multiplying the data by a new standard deviation, we \emph{modulate} the weights:

\begin{equation}
u'_{ijk}=s_i \cdot u_{ijk},
\end{equation}

and instead of normalizing the data, we \emph{demodulate} the weights:

\begin{equation}
u''_{ijk}= u'_{ijk} \bigg/ \sqrt{\sum_{i,k} {u'_{ijk}}^2}.
\end{equation}

\paragraph{Discriminator}
\begin{figure}
\centering
\includegraphics[width=\linewidth]{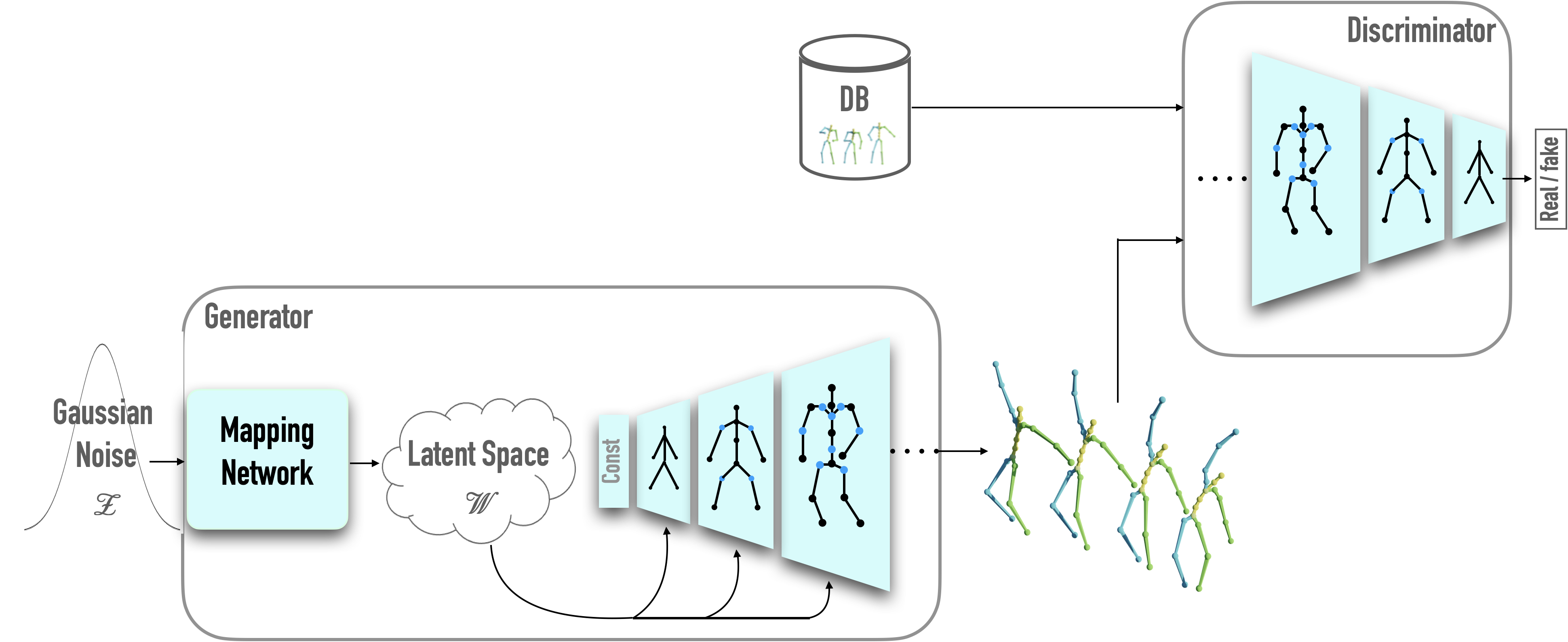} 

\caption{Our discriminator holds the reverse architecture of the synthesis network.
It receives a generated or real motion, and learns to output whether the input motion is real of fake. Using structure-aware neural modules, in each hierarchical level the skeletal topology becomes coarser and the number of frames is divided by 2. }

\label{fig:arch_disc}
\end{figure}

Our discriminator, as well as its role in the training procedure, is depicted in \cref{fig:arch_disc}. Our discriminator holds the reverse architecture of the synthesis network.
That is, it receives a generated or real motion, and processes it in neural blocks that gradually decrease in size. Like the motion synthesis network, our discriminator is based on structure-aware neural modules. In each hierarchical level, the skeletal topology becomes coarser using skeletal convolutions.

\paragraph{Losses} \label{sec:gan_losses}
The generative network is trained with several losses.
Our main loss is adversarial. In addition, we regularize the generator with foot contact and with path length, and regularize the discriminator with $R1$.
All losses except for foot contact are used by StyleGAN too, and for completeness we describe them here.

\subparagraph{Adversarial loss}
We train our GAN with a non-saturating adversarial loss~\cite{goodfellow2014generative},

\begin{equation}
\Loss_{adv}^G =  
- \underset{z \sim \zz}{\E} \left[ \log D(G(z)) \right],
\end{equation}

\begin{eqnarray}
\Loss_{adv}^D  & = & 
-   \underset{m\sim \mm_{nat}}{\E}   \left[ \log D(m) \right] \\
\nonumber & = & \nonumber  -   \underset{z \sim \zz}{\E}   \left[ \log (1-D(G(z))) \right].
\end{eqnarray} 

\subparagraph{Path length loss}
This loss~\cite{karras2020analyzing} requires that a fixed-size step in $\ww$ results in a non-zero, fixed-magnitude change in the generated motion.

\begin{equation}
\Loss_{path}^G  =  
\underset{w\sim\ww,r\sim\rr}{\E} \left[ \norm{J_w^T G(w)*r}_2-a \right]^2,
\end{equation}

where $\rr$ is a unit Gaussian space normalized by the number of joints and frames, $J_w \narroweq \partial G(w)/ \partial(w)$, and $a$ is the accumulated mean gradient length.

\subparagraph{R1 loss}
This loss~\cite{MeschederICML2018} improves the functioning of the discriminator:

\begin{equation}
\Loss_{R1}^D  =  
\underset{m\sim \mm_{nat}}{\E} \left[ \norm{\nabla_m D(m)}_2^2 \right].
\end{equation}

\subparagraph{foot contact loss} The foot contact losses, $\Loss_{tch}^G$ and $\Loss_{fcon}^G$ are described in the main paper.

Altogether, the generator and discriminator losses are 

\begin{eqnarray}
\Loss^G   &=&  \Loss_{adv}^G   +   \lambda_{tch}\Loss_{tch}^G   +   \lambda_{fcon}^G\Loss_{fcon}^G, \\
\Loss^D   &=&   \Loss_{adv}^D.
\end{eqnarray}

We activate the regularizations $\Loss_{path}^G$ and $\Loss_{R1}^D$ in a lazy fashion, as done by Karras \etal \shortcite{karras2020analyzing}.

\subsection{Encoder -- Description of \texorpdfstring{$\ww+$}{TEXT}} \label{sec:w_plus}%

Our inversion method uses the $\ww+$ space, which as an expansion of the latent space $\ww$, proposed by Abdal \etal \shortcite{abdal2019image2stylegan}. A tensor in $\ww+$ is a concatenation of several different $w\in\ww$ vectors, one for each layer of the synthesis network. Each vector in $\ww+$ is used as a modulation input to a different layer. In contrast, when using $\ww$, the same $w$ vector is used for all layers. Abdal \etal \shortcite{abdal2019image2stylegan} show that $\ww$ is limited and an inversion from arbitrary images is much more accurate when using $\ww+$. In our experiments we have witnessed that this approach works for the motion domain as well.

\section{Applications -- Additional Details } \label{sec:app_details}

\subsection{Latent interpolation -- Additional Details} \label{sec:latent_interp}

\begin{figure}
\centering
\includegraphics[width=\linewidth]{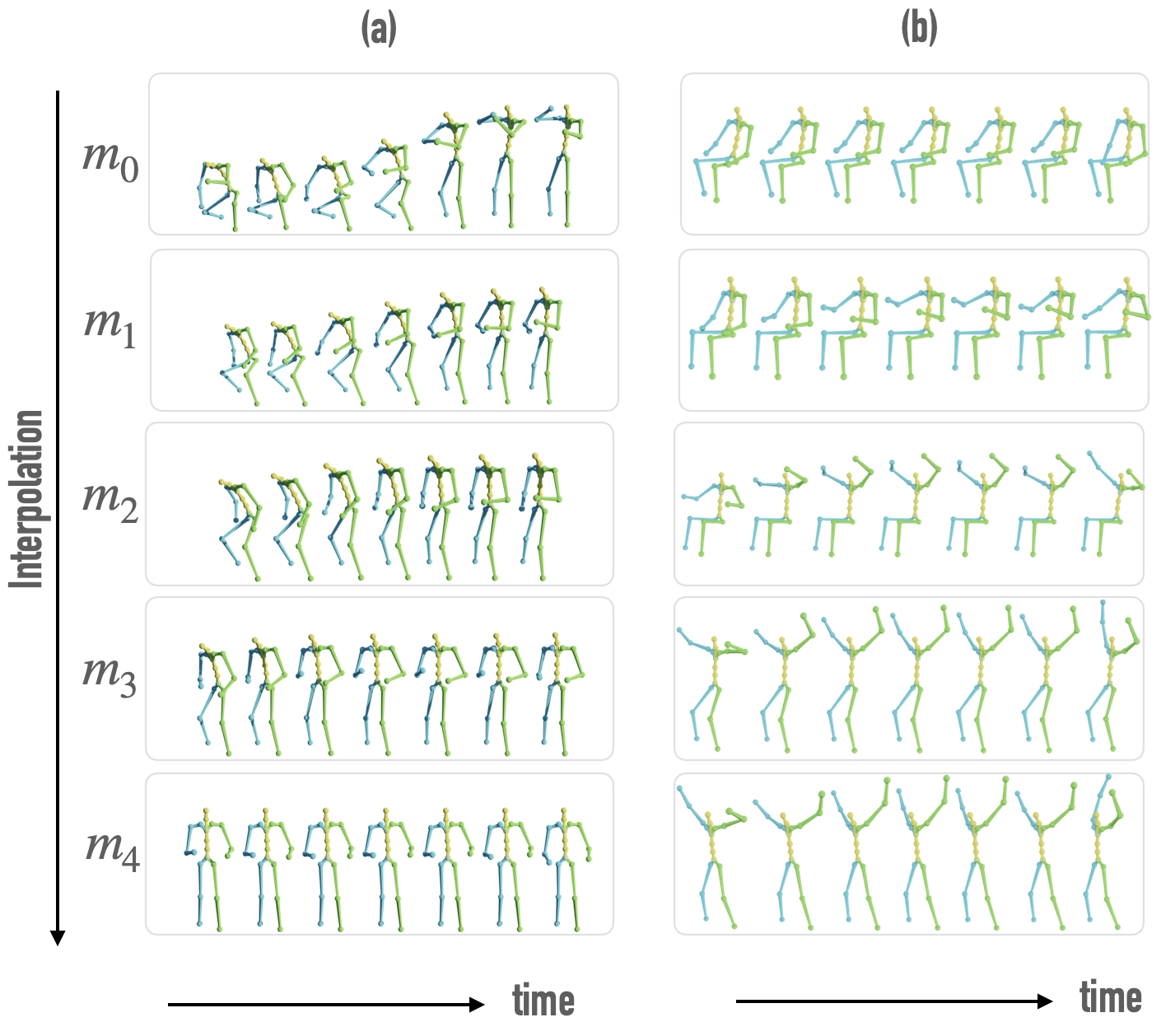} 

\setlength{\abovecaptionskip}{0pt plus 3pt minus 2pt}
\setlength{\belowcaptionskip}{-10pt plus 3pt minus 2pt}

\caption{
Interpolation in the latent space: 
$(a)$ interpolation to the mean motion (truncation); $(b)$ interpolation between sampled motions. Note that the interpolated motions seem natural, even if the interpolation is performed between sitting and standing, a results that can be achieved if the interpolation is performed between the joint positions.
}
\label{fig:interp}

\end{figure}

In this section we elaborate regarding interpolation in the latent space, referred in the main paper.
\Cref{fig:interp}, also shown in the main paper, is copied here so we can further describe it.

Let $\mean{w}$ be the mean of all $w\in\ww$, and let \emph{mean motion} denote $G(\mean{w})$, the motion generated by it. 
The mean motion is depicted at the bottom row of  \cref{fig:interp}(a). This motion is similar for all variations of trained networks, and is what we  intuitively expect: an idle standing, front facing character.

We demonstrate the linearity of the latent space $\ww$ by
interpolating between the latent values 
and observing the motions generated out of the interpolated values.
A special case, called \emph{truncation}, is when the interpolation target is $\mean{w}$. 
In the imaging domain, truncation has an important role in regularizing out of distribution images. We show that truncation works well in our model too.
A truncated sequence is denoted by $w_i=\hat{w}+\frac{i}{C}(\mean{w}-\hat{w})$, where $\hat{w}\in \ww$, $C$ is the number of interpolation steps, and $i\in [0\dots C ]$. Clearly $w_0=\hat{w}$ and $w_{C}=\mean{w}$.
We can replace $\mean{w}$ by any sampled $\tilde{w}\in\ww$, and then the sequence is called interpolated rather than truncated.
Let $m_i=G(w_i)$ denote the motion generated out of each $w_i$.
\cref{fig:interp} (a) and (b) shows the motions create out of truncation and interpolation,
respectively.

We observe favorable characteristics in all interpolation sequences.
First, $m_i$ is semantically similar to $m_{i-1}$, but it also changed towards the semantics of the target $m_C$.
When dealing with truncation, $m_i$ is always milder than $m_{i-1}$.

Second, we notice that the interpolation is 
between whole sequences rather than frames.
For example, if in $m_{i-1}$ the character jumps occasionally, then in $m_i$ the character jumps in a similar frequency, but unnecessarily on the same frames.  

Lastly, there are no unnatural motions in the sequence, although 
using simple geometric joint interpolation would have resulted in unnatural motions.
\cref{fig:interp}(b) demonstrates this, where our latent interpolation yields natural motions at all stages. 
A n\"aive geometric interpolation of edge rotations would result in abnormal pose between sitting to standing, with a vertical spine (see supplementary video).

\subsection{Computing the \emph{gral} score} \label{sec:gral_score}
Our classifier computes the \emph{gral} (gradual right arm lifting) score in the following way. Let $m=[R,S,F]$ be a selected motion. Recall $R$ represents the rotation angles of the motion. Let $R_{rs,t}$ and $R_{re,t}$ denote the rotations of the right shoulder and the right elbow at time $t$, respectively.
Let $[R_{rs,t},...,R_{rs,t+8}]$ be a temporal window of size 8. A similar window is created for $R_{re}$. We compute the average angle in each window, and slide the window with stride of 4. Altogether we get the average computed $T/4$ times for both the right shoulder and the right elbow. Denote the sequence of average angles by $\alpha_{rs}$ and $\alpha_{re}$. The next step is to compute the difference between each element to the one preceding it, and obtain 

\begin{equation}
\smallmath
score_{rs_i}=
    \begin{cases}
      1, & \text{if}\ \alpha_{rs_i}>\alpha_{rs_{i-1}} \\
      0, & \text{otherwise}
    \end{cases},
\end{equation}
\begin{equation}
\smallmath
score_{re_i}=
    \begin{cases}
      1, & \text{if}\ \alpha_{re_i}>\alpha_{re_{i-1}} \\
      0, & \text{otherwise}
    \end{cases},
\end{equation}

where $i\in [1,T/4-1]$.
   
Clearly, if all scores are one, the arm is going up, and if they are all zero, the arm is going down. The average of all the values in the two score vectors is used as the final attribute score.

\begin{figure*}[th]
\centering
\includegraphics[width=\linewidth]{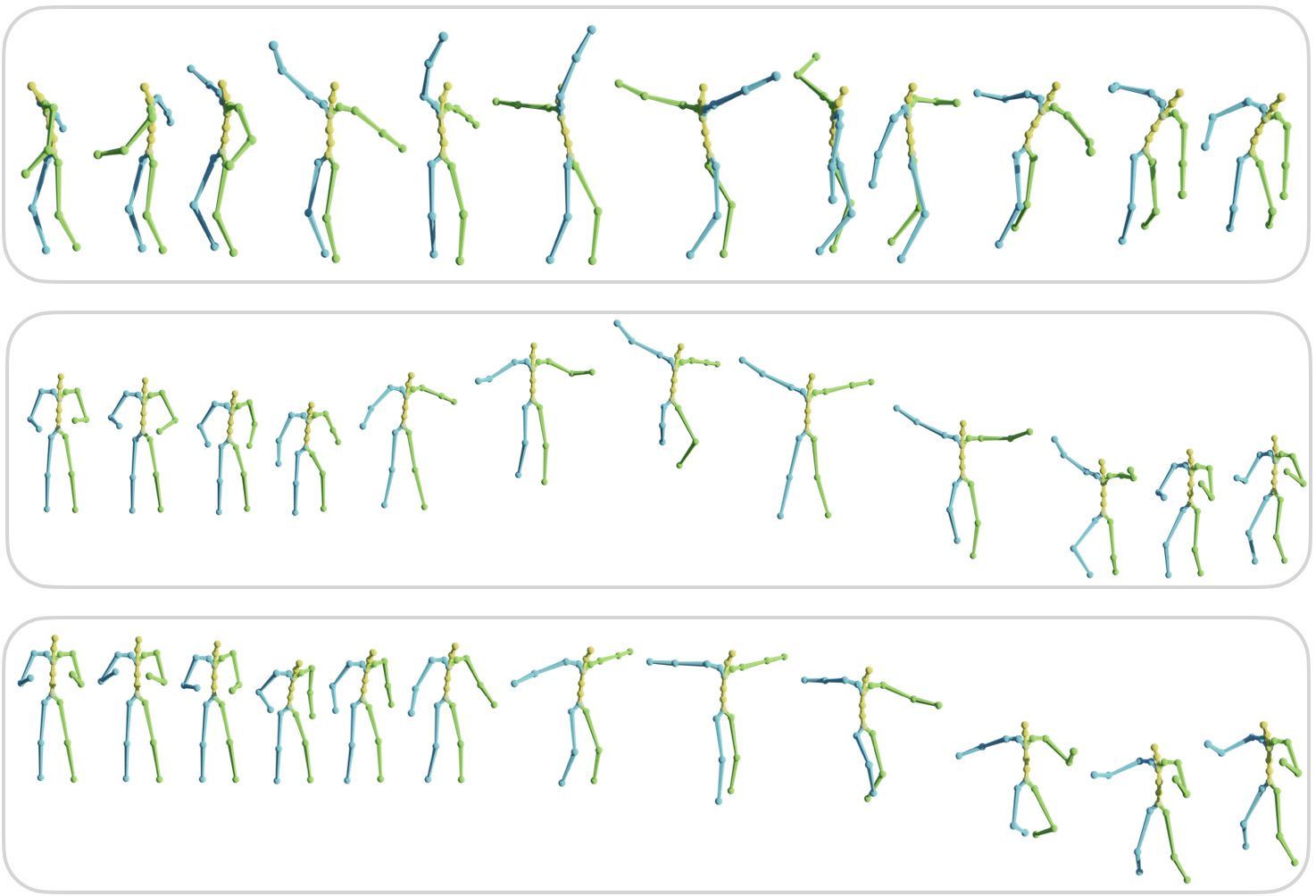}

\caption{Qualitative results. The top motion depicts a synthesised wild dance, and the next two show synthesised jumps, illustrating the diversity in semantically related motions. All motions are unconditionally generated. See more results in the supplementary video.}
\label{fig:qual_sup}
\end{figure*}

\section{Experiments -- Additional Details} \label{sec:experiments_detail}
\subsection{Datasets} \label{sec:datasets}
\paragraph{Mixamo -- training and evaluation} 
We construct our 3D motion dataset using the Mixamo \cite{mixamo} 3D animation collection, which contains approximately 2500 extremely diverse motions that are not constrained by any set of categories. These motions are applied on 70 characters.
Examples of the motions in the dataset are elementary actions (jumping, walking), dance moves (samba, hip-hop), martial arts (boxing, capoeira), acrobatics (back/front flips, acrobatic jumps), and non-standard motions (running on a wall, flying). 

We generate our data samples by first extracting the relevant edges from each motion (\eg, we drop the fingers). Then we crop each motion to partially overlapping sequences of frames, hence increasing the amount of data.

\paragraph{HumanAct12 -- evaluation}
HumanAct12~\cite{guo2020action2motion} is not as diverse as Mixamo, and offers approximately 1200 motion clips, organized into 12 action categories and 34 subcategories. Due to its small number of motions, we use HumanAct12 for quantitative comparison only.

\subsection{Hyper-parameters and Training Details} \label{sec:hyperparameters}
\begin{table}
\small
\centering
    \begin{tabular}{ c c c}
    \toprule
    Name & \makecell{Hierarchy \\ level}  & channels $\times$ joints $\times$ frames\\
    \toprule 
    Generator - 
     & 0  &  256 $\times$ 1 $\times$ 4                    \\
     Motion Synth. Net.
     & 1  &  128 $\times$ 2 $\times$ 8                    \\
     & 2  &  64 $\times$ 7 $\times$ 16                    \\
     & 3  &  64 $\times$ 12 $\times$ 32                     \\
     & 4  &  32 $\times$ 20 $\times$ 64                     \\
    \midrule Discriminator
     & 0  &  32 $\times$ 20 $\times$ 64                    \\
     & 1  &  64 $\times$ 12 $\times$ 32                     \\
     & 2  &  64 $\times$ 7 $\times$ 16                      \\
     & 3  &  128 $\times$ 2 $\times$ 8                      \\
     & 4  &   256 $\times$ 1 $\times$ 4                      \\
    \bottomrule
    \end{tabular}
\caption{Architecture: Dimensions of all hierarchy levels.}
\label{tab:tech_sizes}
\end{table}

\begin{table}
\small
\centering
    \begin{tabular}{ c c}
    \toprule
    Name & Neural building blocks\\
    \toprule 
    Generator -
     & Skeletal Conv. Scaler (upsample)      \\
    Motion Synth. Net.
     & Skeletal Conv. (in-place)                     \\
     & Skeletal Convl (in-place)                    \\
    \midrule Discriminator 
     & Skeletal Conv. (in-place)                    \\
     & Skeletal Conv. (in-place)                    \\
     & Skeletal Conv. Scaler (downsample)         \\
     & Add Residual \\
    \bottomrule
    \end{tabular}
\caption{Architecture: Building blocks in hierarchical levels. Skeletal operators are based on ~\cite{aberman2020skeleton}.}

\label{tab:tech_architecture}
\end{table}

In this section, we describe the details for the network architectures.
\cref{tab:tech_sizes} describes the architecture of our generator and discriminator networks. The sizes of the kernels are configurable by hyper-parameters, and in the table we specify which hyper-parameters we have used for our best model. Note that the number of joints varies according to the topology of the skeleton on which the network is trained. The values in \cref{tab:tech_sizes} belong to the skeleton used by the model presented in this work. 
The structure of each hierarchical level in our generator and discriminator is described in \cref{tab:tech_architecture}. A hierarchy level in the motion synthesis network contains input/output skips, and a hierarchy level in the discriminator contains a residual skip, both based on Karras \etal \shortcite{karras2020analyzing}.

In our experiments, we use $\lambda_{fcon}=1$, $\lambda_{touch}=0.01$, batch size 16, learning rate 0.002 for both generator and discriminator, mixing 0.9, and train for 80,000 iterations. We use pytorch version 1.5.0, and CUDA version 10.1 on a GeForce GTX 1080 Ti GPU.

\subsection{Quantitative Metrics} \label{sec:metrics}

Our metrics build upon the latent features of an action recognition model.
However, training such a model on Mixamo is challenging, as there are no action labels in it.  

Our approach to this challenge is interdisciplinary. Mixamo has textual labels, and using the Sentence-BERT~\cite{reimers2019sentence} NLP model, we attain latent features representing the textual characteristics of each motion. Then we use K-means to cluster the embedding, 
and use each cluster as a pseudo action label.
With action labels at hand, we train an action recognition model -- STGCN~\cite{yan2018spatial}. The features extracted from this trained model are then used for metrics calculation. 
We randomly sample 2000 motions for calculating metric scores on the Mixamo dataset. We draw 1000 motions for scores on the HumanAct12 dataset, since it is a lot smaller.

Following is a brief description for each metric used for the quantitative results.
\paragraph{FID}
Fr\`{e}chet inception distance is the distance between the feature distribution of generated motions and that of the real motions, namely the difference in mean and variance. Despite its simplicity, FID is an important metric widely used to evaluate the overall quality of generated motions~\cite{guo2020action2motion, petrovich2021action}. 
FID is borrowed from the image domain, where  the inception network is used for features. 
To adjust this metric to the motion domain, we replace the inception by an action recognition network. A lower value implies better FID results.

\paragraph{KID}
Kernel Inception Distance (KID), proposed by Binkowski \etal \shortcite{binkowski2018demystifying}, compares skewness as well as the values compared in FID, namely mean and variance. KID is known to work better for small and medium size datasets. Lower values are better.

\paragraph{Precision and Recall}
These measures are adopted from the discriminative domain to the generative domain~\cite{sajjadi2018assessing}. Precision measures the probability that a randomly generated motion falls within the support of the distribution of real images, and is closely related with fidelity. Recall measures the probability that a real motion falls within the support of the distribution of generated images, and is closely related with diversity. Higher precision and recall values imply better results.

\paragraph{Diversity}
This metric measures the variance of generated motions~\cite{guo2020action2motion, petrovich2021action}. In the context of  action recognition models, it measures the variance across all action categories, and therefore it suits an unconstrained generator. The diversity value is considered good if it is close to the diversity of the ground truth. 
In all our experiments, the diversity of the generated data was lower than the ground truth, so for clarity we mark it with an upwards pointing arrow, implying that for our case, higher is better.

\sr{
\subsection{Additional Ablation} \label{sec:more_ablation}
\begin{table}[t] 

\centering

\resizebox{\columnwidth}{!}{

\begin{tabular}{c|*{5}{c}}
\toprule
\textbf{\diagbox{Loss}{Error}}  & \textbf{\makecell{Reconst.\\ (L2)}} & \textbf{\makecell{Reconst.\\ (L1)}} & \textbf{\makecell{Global root \\position (mm)}} & \textbf{\makecell{Local \\position (mm)}} & \textbf{\makecell{Global \\position (mm)}} \\
\midrule
\textbf{all}                  & $\underline{.293}$ & $\underline{.316}$ & $59.0$             & $\textbf{20.4}$       & $\underline{78.2}$ \\
\midrule
\boldmath\textbf{w/o $\Loss_{fcon}^I$} & $\textbf{.271}$       & $\textbf{.293}$       & $\underline{57.0}$ & $\underline{21.2}$ & $\textbf{76.0}$       \\
\midrule
\boldmath\textbf{w/o $\Loss_{pos}^I$}  & $.300$             & $.321$             & $\textbf{47.0}$       & $47.7$             & $93.7$             \\
\midrule
\boldmath\textbf{w/o} $\Loss_{root}^I$ & $.319$            & $.328$              & $472.2$            & $27.3$             & $493.5$            \\
\bottomrule
\end{tabular}

} %

\caption{Quantitative results for the encoder losses, on the Mixamo encoder test set. 
Best scores are emphasised in \textbf{bold}, second best are \underline{underlined}.
}
\label{tab:ablation_enc_loss}

\end{table}

In \cref{tab:ablation_enc_loss} we conduct an ablation study of the encoder losses. 
The best scores are mostly obtained when \emph{not} using the foot contact loss, and the second best ones are mostly obtained when using all losses. This is expected, as the foot contact loss biases the results towards more accurate foot contact on the account of other body parts accuracy. However, qualitatively, the human eye prefers coherent foot contact, and in our supplementary video the pleasing foot contact results can be noticed. The phenomenon of foot contact loss degrading the quantitative results, but upgrading the qualitative ones, has been also reported in a concurrent work, MDM~\cite{tevet2022human}.

As detailed in the main paper, the losses of the encoder are a reconstruction loss $\Loss_{rec}^I$, a foot contact loss $\Loss_{fcon}^I$, a root loss $\Loss_{root}^I$, and a position loss $\Loss_{pos}^I$. We measure the performance of the encoder using several metrics. 
Recall $\mm_{tst}$ denotes the encoder test set, $m$ denotes an unseen motion, $I$ denotes our trained encoder and $G$ denotes our trained generator.

\paragraph{Reconstruction L2 Error}
This is the most important metric, as it makes sure the encoder is fulfilling its goal, \ie project a motion data structure into the latent space such that the motion data structure generated from the projected value is as similar as possible to the original one. This metric is identical to the reconstruction loss, $\Loss_{rec}^I$, and is measured with 
\begin{equation}
E_{recL2}^I  =  \underset{m\sim \mm_{tst}}{\E} \left[ \norm{m-G(I(m))}_2^2\right]. 
\end{equation}

\paragraph{Reconstruction L1 Error} 
Same as the previous metric, but this time with L1. The error is measured by
\begin{equation}
E_{recL1}^I  =  \underset{m\sim \mm_{tst}}{\E} \left[ \norm{m-G(I(m))}_1\right]. 
\end{equation}

\paragraph{Position Error} 
In addition to the reconstruction error that mainly measures rotation angle error, we measure the error of the joint position themselves. Since the global root position has a large error component, we split the error into the global root position error only, local position error relative to the root, and both accumulated together. These errors are measured by 
\begin{equation}
E_{rt}^I  \narroweq  \underset{m\sim \mm_{tst}}{\E} \left[ \norm{FK(m)_{rt} \narrowmin FK(G(I(m)))_{rt}}_2^2\right], 
\end{equation}
\begin{equation}
E_{nrt}^I  \narroweq\!\!\!\!  \underset{m\sim \mm_{tst}}{\E}\! \left[\! \norm{FK(m)_{nrt} \!\narrowmin FK(G(I(m))_{nrt})}_2^2\right], 
\end{equation}
\begin{equation}
E_{pos}^I  \narroweq  \underset{m\sim \mm_{tst}}{\E} \left[ \norm{FK(m) \narrowmin FK(G(I(m)))}_2^2\right], 
\end{equation}
where $FK$ is a forward kinematic operator yielding joint locations, $(\cdot)_{rt}$ is the root component of the position array, and $(\cdot)_{nrt}$ is the position array excluding its root component.

\begin{table}[t] 

\centering

\resizebox{\columnwidth}{!}{

\begin{tabular}{c|*{5}{c}}
\toprule
\textbf{\diagbox{Repr.}{Metric}}  & \textbf{FID $\downarrow$} & \textbf{KID $\downarrow$} & \pmb{Diversity $\uparrow$} & \textbf{Precision $\uparrow$} & \textbf{Recall $\uparrow$} \\
\midrule
\textbf{Velocity} & $11.3$       & $.118$       & $\textbf{15.8}$ & $\textbf{.470}$ & $\textbf{.696}$ \\
\midrule
\textbf{Location} & $\textbf{10.7}$ & $\textbf{.113}$ & $15.1$       & $.468$       & $.695$       \\
\bottomrule
\end{tabular}

} %

\caption{Quantitative results for the root position representation, on the Mixamo dataset. 
Best scores are emphasised in \textbf{bold}.
}
\label{tab:ablation_pos_vs_velo}

\end{table}

In \cref{tab:ablation_pos_vs_velo} we run an ablation study of the root position representation. Predicting a root position that faithfully reflects the dataset and yields natural motions is challenging, and many existing works either avoid predicting global position, or predict it inaccurately, yielding a floating or jittery appearance. We study two possible representations for the root; a 3D location, or its velocity. The quantitative results of the two representations are comparable, and yet, the qualitative results have been in favor of the velocity representation, hence our choice.
}

\subsection{Additional Qualitative Results} \label{sec:qual_sup}
In \cref{fig:qual_sup} we show additional qualitative results. The reader is encouraged to watch the supplementary video in order to get the full impression of our results.

\fi

\ifarxiv
\clearpage
\fi

{\small
\bibliographystyle{CVPR/ieee_fullname}
\bibliography{main}

\begin{thebibliography}{10}\itemsep=-1pt

\bibitem{abdal2019image2stylegan}
Rameen Abdal, Yipeng Qin, and Peter Wonka.
\newblock Image2stylegan: How to embed images into the stylegan latent space?
\newblock In {\em Proceedings of the IEEE/CVF International Conference on
  Computer Vision}, pages 4432--4441, 2019.

\bibitem{aberman2020skeleton}
Kfir Aberman, Peizhuo Li, Dani Lischinski, Olga Sorkine-Hornung, Daniel
  Cohen-Or, and Baoquan Chen.
\newblock Skeleton-aware networks for deep motion retargeting.
\newblock {\em ACM Transactions on Graphics (TOG)}, 39(4):62--1, 2020.

\bibitem{aberman2020unpaired}
Kfir Aberman, Yijia Weng, Dani Lischinski, Daniel Cohen-Or, and Baoquan Chen.
\newblock Unpaired motion style transfer from video to animation.
\newblock {\em ACM Transactions on Graphics (TOG)}, 39(4):64--1, 2020.

\bibitem{aberman2019learning}
Kfir Aberman, Rundi Wu, Dani Lischinski, Baoquan Chen, and Daniel Cohen-Or.
\newblock Learning character-agnostic motion for motion retargeting in 2d.
\newblock {\em arXiv preprint arXiv:1905.01680}, 2019.

\bibitem{mixamo}
{Adobe Systems Inc.}
\newblock Mixamo, 2021.
\newblock Accessed: 2021-12-25.

\bibitem{ahuja2019language2pose}
Chaitanya Ahuja and Louis-Philippe Morency.
\newblock Language2pose: Natural language grounded pose forecasting.
\newblock In {\em 2019 International Conference on 3D Vision (3DV)}, pages
  719--728. IEEE, 2019.

\bibitem{aksan2019structured}
Emre Aksan, Manuel Kaufmann, and Otmar Hilliges.
\newblock Structured prediction helps 3d human motion modelling.
\newblock In {\em Proceedings of the IEEE/CVF International Conference on
  Computer Vision}, pages 7144--7153. IEEE Computer Society, 2019.

\bibitem{aristidou2021rhythm}
Andreas Aristidou, Anastasios Yiannakidis, Kfir Aberman, Daniel Cohen-Or, Ariel
  Shamir, and Yiorgos Chrysanthou.
\newblock Rhythm is a dancer: Music-driven motion synthesis with global
  structure.
\newblock {\em arXiv preprint arXiv:2111.12159}, 2021.

\bibitem{barsoum2018hp}
Emad Barsoum, John Kender, and Zicheng Liu.
\newblock Hp-gan: Probabilistic 3d human motion prediction via gan.
\newblock In {\em Proceedings of the IEEE conference on computer vision and
  pattern recognition workshops}, pages 1418--1427, 2018.

\bibitem{bermano2022state}
Amit~H Bermano, Rinon Gal, Yuval Alaluf, Ron Mokady, Yotam Nitzan, Omer Tov,
  Oren Patashnik, and Daniel Cohen-Or.
\newblock State-of-the-art in the architecture, methods and applications of
  stylegan.
\newblock In {\em Computer Graphics Forum}, volume~41, pages 591--611. Wiley
  Online Library, 2022.

\bibitem{bhattacharya2021text2gestures}
Uttaran Bhattacharya, Nicholas Rewkowski, Abhishek Banerjee, Pooja Guhan,
  Aniket Bera, and Dinesh Manocha.
\newblock Text2gestures: A transformer-based network for generating emotive
  body gestures for virtual agents.
\newblock In {\em 2021 IEEE Virtual Reality and 3D User Interfaces (VR)}, pages
  1--10. IEEE, 2021.

\bibitem{biswas2021hierarchical}
Sourav Biswas, Kangxue Yin, Maria Shugrina, Sanja Fidler, and Sameh Khamis.
\newblock Hierarchical neural implicit pose network for animation and motion
  retargeting.
\newblock {\em arXiv preprint arXiv:2112.00958}, 2021.

\bibitem{binkowski2018demystifying}
Mikołaj Bińkowski, Dougal~J. Sutherland, Michael Arbel, and Arthur Gretton.
\newblock Demystifying {MMD} {GAN}s.
\newblock In {\em International Conference on Learning Representations}, 2018.

\bibitem{brock2018large}
Andrew Brock, Jeff Donahue, and Karen Simonyan.
\newblock Large scale gan training for high fidelity natural image synthesis.
\newblock {\em arXiv preprint arXiv:1809.11096}, 2018.

\bibitem{cervantes2022implicit}
Pablo Cervantes, Yusuke Sekikawa, Ikuro Sato, and Koichi Shinoda.
\newblock Implicit neural representations for variable length human motion
  generation.
\newblock {\em arXiv preprint arXiv:2203.13694}, 2022.

\bibitem{degardin2022generative}
Bruno Degardin, Jo{\~a}o Neves, Vasco Lopes, Jo{\~a}o Brito, Ehsan Yaghoubi,
  and Hugo Proen{\c{c}}a.
\newblock Generative adversarial graph convolutional networks for human action
  synthesis.
\newblock In {\em Proceedings of the IEEE/CVF Winter Conference on Applications
  of Computer Vision}, pages 1150--1159, Los Alamitos, CA, USA, 2022. IEEE
  Computer Society.

\bibitem{duan2021single}
Yinglin Duan, Tianyang Shi, Zhengxia Zou, Yenan Lin, Zhehui Qian, Bohan Zhang,
  and Yi Yuan.
\newblock Single-shot motion completion with transformer.
\newblock {\em arXiv preprint arXiv:2103.00776}, 2021.

\bibitem{fragkiadaki2015recurrent}
Katerina Fragkiadaki, Sergey Levine, Panna Felsen, and Jitendra Malik.
\newblock Recurrent network models for human dynamics.
\newblock In {\em Proceedings of the IEEE International Conference on Computer
  Vision}, pages 4346--4354. IEEE Computer Society, 2015.

\bibitem{ghorbani2020b}
Saeed Ghorbani, Calden Wloka, Ali Etemad, Marcus~A. Brubaker, and Nikolaus~F.
  Troje.
\newblock Probabilistic character motion synthesis using a hierarchical deep
  latent variable model.
\newblock {\em Computer Graphics Forum}, 2020.

\bibitem{goodfellow2014generative}
Ian Goodfellow, Jean Pouget-Abadie, Mehdi Mirza, Bing Xu, David Warde-Farley,
  Sherjil Ozair, Aaron Courville, and Yoshua Bengio.
\newblock Generative adversarial nets.
\newblock {\em Advances in neural information processing systems}, 27, 2014.

\bibitem{gordon2022flex}
Brian Gordon, Sigal Raab, Guy Azov, Raja Giryes, and Daniel Cohen-Or.
\newblock Flex: Extrinsic parameters-free multi-view 3d human motion
  reconstruction.
\newblock In {\em European Conference on Computer Vision (ECCV)}, pages
  176--196. Springer, 2022.

\bibitem{guo2022generating}
Chuan Guo, Shihao Zou, Xinxin Zuo, Sen Wang, Wei Ji, Xingyu Li, and Li Cheng.
\newblock Generating diverse and natural 3d human motions from text.
\newblock In {\em Proceedings of the IEEE/CVF Conference on Computer Vision and
  Pattern Recognition}, pages 5152--5161, 2022.

\bibitem{guo2020action2motion}
Chuan Guo, Xinxin Zuo, Sen Wang, Shihao Zou, Qingyao Sun, Annan Deng, Minglun
  Gong, and Li Cheng.
\newblock Action2motion: Conditioned generation of 3d human motions.
\newblock In {\em Proceedings of the 28th ACM International Conference on
  Multimedia}, pages 2021--2029. ACM New York, NY, USA, 2020.

\bibitem{habibie2017recurrent}
Ikhsanul Habibie, Daniel Holden, Jonathan Schwarz, Joe Yearsley, and Taku
  Komura.
\newblock A recurrent variational autoencoder for human motion synthesis.
\newblock In {\em 28th British Machine Vision Conference}, 2017.

\bibitem{harvey2018recurrent}
F{\'e}lix~G Harvey and Christopher Pal.
\newblock Recurrent transition networks for character locomotion.
\newblock In {\em SIGGRAPH Asia 2018 Technical Briefs}, pages 1--4, 2018.

\bibitem{harvey2020robust}
F{\'e}lix~G Harvey, Mike Yurick, Derek Nowrouzezahrai, and Christopher Pal.
\newblock Robust motion in-betweening.
\newblock {\em ACM Transactions on Graphics (TOG)}, 39(4):60--1, 2020.

\bibitem{hernandez2019human}
Alejandro Hernandez, Jurgen Gall, and Francesc Moreno-Noguer.
\newblock Human motion prediction via spatio-temporal inpainting.
\newblock In {\em Proceedings of the IEEE/CVF International Conference on
  Computer Vision}, pages 7134--7143. IEEE Computer Society, 2019.

\bibitem{ho2020denoising}
Jonathan Ho, Ajay Jain, and Pieter Abbeel.
\newblock Denoising diffusion probabilistic models.
\newblock {\em Advances in Neural Information Processing Systems},
  33:6840--6851, 2020.

\bibitem{holden2017fast}
Daniel Holden, Ikhsanul Habibie, Ikuo Kusajima, and Taku Komura.
\newblock Fast neural style transfer for motion data.
\newblock {\em IEEE computer graphics and applications}, 37(4):42--49, 2017.

\bibitem{holden2017phase}
Daniel Holden, Taku Komura, and Jun Saito.
\newblock Phase-functioned neural networks for character control.
\newblock {\em ACM Transactions on Graphics (TOG)}, 36(4):1--13, 2017.

\bibitem{holden2016deep}
Daniel Holden, Jun Saito, and Taku Komura.
\newblock A deep learning framework for character motion synthesis and editing.
\newblock {\em ACM Transactions on Graphics (TOG)}, 35(4):1--11, 2016.

\bibitem{holden2015learning}
Daniel Holden, Jun Saito, Taku Komura, and Thomas Joyce.
\newblock Learning motion manifolds with convolutional autoencoders.
\newblock In {\em SIGGRAPH Asia 2015 technical briefs}, pages 1--4, 2015.

\bibitem{huang2017arbitrary}
Xun Huang and Serge Belongie.
\newblock Arbitrary style transfer in real-time with adaptive instance
  normalization.
\newblock In {\em Proceedings of the IEEE international conference on computer
  vision}, pages 1501--1510, 2017.

\bibitem{Jang2020constructing}
Deok-Kyeong Jang and Sung-Hee Lee.
\newblock Constructing human motion manifold with sequential networks.
\newblock In {\em Computer Graphics Forum}, volume~39, pages 314--324. Wiley
  Online Library, 2020.

\bibitem{karras2021alias}
Tero Karras, Miika Aittala, Samuli Laine, Erik H{\"a}rk{\"o}nen, Janne
  Hellsten, Jaakko Lehtinen, and Timo Aila.
\newblock Alias-free generative adversarial networks.
\newblock {\em Advances in Neural Information Processing Systems}, 34, 2021.

\bibitem{karras2020analyzing}
Tero Karras, Samuli Laine, Miika Aittala, Janne Hellsten, Jaakko Lehtinen, and
  Timo Aila.
\newblock Analyzing and improving the image quality of stylegan.
\newblock In {\em Proceedings of the IEEE/CVF Conference on Computer Vision and
  Pattern Recognition}, pages 8110--8119, 2020.

\bibitem{kaufmann2020convolutional}
Manuel Kaufmann, Emre Aksan, Jie Song, Fabrizio Pece, Remo Ziegler, and Otmar
  Hilliges.
\newblock Convolutional autoencoders for human motion infilling.
\newblock In {\em 2020 International Conference on 3D Vision (3DV)}, pages
  918--927. IEEE, 2020.

\bibitem{kingma2018glow}
Durk~P Kingma and Prafulla Dhariwal.
\newblock Glow: Generative flow with invertible 1x1 convolutions.
\newblock {\em Advances in neural information processing systems}, 31, 2018.

\bibitem{lee2018listen}
Juheon Lee, Seohyun Kim, and Kyogu Lee.
\newblock Listen to dance: Music-driven choreography generation using
  autoregressive encoder-decoder network.
\newblock {\em arXiv preprint arXiv:1811.00818}, 2018.

\bibitem{li2022ganimator}
Peizhuo Li, Kfir Aberman, Zihan Zhang, Rana Hanocka, and Olga Sorkine-Hornung.
\newblock Ganimator: Neural motion synthesis from a single sequence.
\newblock {\em ACM Transactions on Graphics (TOG)}, 41(4):138, 2022.

\bibitem{li2021learn}
Ruilong Li, Shan Yang, David~A Ross, and Angjoo Kanazawa.
\newblock Learn to dance with aist++: Music conditioned 3d dance generation.
\newblock {\em arXiv e-prints}, pages arXiv--2101, 2021.

\bibitem{maheshwari2022mugl}
Shubh Maheshwari, Debtanu Gupta, and Ravi~Kiran Sarvadevabhatla.
\newblock Mugl: Large scale multi person conditional action generation with
  locomotion.
\newblock In {\em Proceedings of the IEEE/CVF Winter Conference on Applications
  of Computer Vision}, pages 257--265, Los Alamitos, CA, USA, 2022. IEEE
  Computer Society.

\bibitem{MeschederICML2018}
Lars Mescheder, Andreas Geiger, and Sebastian Nowozin.
\newblock Which training methods for gans do actually converge?
\newblock In {\em International Conference on Machine learning (ICML)}, 2018.

\bibitem{nitzan2021large}
Yotam Nitzan, Rinon Gal, Ofir Brenner, and Daniel Cohen-Or.
\newblock Large: Latent-based regression through gan semantics.
\newblock {\em arXiv preprint arXiv:2107.11186}, 2021.

\bibitem{petrovich2021action}
Mathis Petrovich, Michael~J Black, and G{\"u}l Varol.
\newblock Action-conditioned 3d human motion synthesis with transformer vae.
\newblock In {\em Proceedings of the IEEE/CVF International Conference on
  Computer Vision}, pages 10985--10995. IEEE Computer Society, 2021.

\bibitem{Petrovich2022temos}
Mathis Petrovich, Michael~J. Black, and Gül Varol.
\newblock Temos: Generating diverse human motions from textual descriptions,
  2022.

\bibitem{reimers2019sentence}
Nils Reimers and Iryna Gurevych.
\newblock Sentence-bert: Sentence embeddings using siamese bert-networks.
\newblock In {\em Proceedings of the 2019 Conference on Empirical Methods in
  Natural Language Processing and the 9th International Joint Conference on
  Natural Language Processing (EMNLP-IJCNLP)}, pages 3982--3992, 2019.

\bibitem{richardson2021encoding}
Elad Richardson, Yuval Alaluf, Or Patashnik, Yotam Nitzan, Yaniv Azar, Stav
  Shapiro, and Daniel Cohen-Or.
\newblock Encoding in style: a stylegan encoder for image-to-image translation.
\newblock In {\em Proceedings of the IEEE/CVF conference on computer vision and
  pattern recognition}, pages 2287--2296, 2021.

\bibitem{sajjadi2018assessing}
Mehdi~SM Sajjadi, Olivier Bachem, Mario Lucic, Olivier Bousquet, and Sylvain
  Gelly.
\newblock Assessing generative models via precision and recall.
\newblock {\em Advances in Neural Information Processing Systems}, 31, 2018.

\bibitem{shen2020interfacegan}
Yujun Shen, Ceyuan Yang, Xiaoou Tang, and Bolei Zhou.
\newblock {InterFaceGAN:} interpreting the disentangled face representation
  learned by {GANs}.
\newblock {\em arXiv preprint arXiv:2005.09635}, 2020.

\bibitem{shi2020motionet}
Mingyi Shi, Kfir Aberman, Andreas Aristidou, Taku Komura, Dani Lischinski,
  Daniel Cohen-Or, and Baoquan Chen.
\newblock Motionet: 3d human motion reconstruction from monocular video with
  skeleton consistency.
\newblock {\em ACM Transactions on Graphics (TOG)}, 40(1):1--15, 2020.

\bibitem{sun2020deepdance}
Guofei Sun, Yongkang Wong, Zhiyong Cheng, Mohan~S Kankanhalli, Weidong Geng,
  and Xiangdong Li.
\newblock Deepdance: music-to-dance motion choreography with adversarial
  learning.
\newblock {\em IEEE Transactions on Multimedia}, 23:497--509, 2020.

\bibitem{tevet2022motionclip}
Guy Tevet, Brian Gordon, Amir Hertz, Amit~H Bermano, and Daniel Cohen-Or.
\newblock Motionclip: Exposing human motion generation to clip space.
\newblock {\em arXiv preprint arXiv:2203.08063}, 2022.

\bibitem{tevet2022human}
Guy Tevet, Sigal Raab, Brian Gordon, Yonatan Shafir, Amit~H Bermano, and Daniel
  Cohen-Or.
\newblock Human motion diffusion model.
\newblock {\em arXiv preprint arXiv:2209.14916}, 2022.

\bibitem{villegas2018neural}
Ruben Villegas, Jimei Yang, Duygu Ceylan, and Honglak Lee.
\newblock Neural kinematic networks for unsupervised motion retargetting.
\newblock In {\em Proceedings of the IEEE Conference on Computer Vision and
  Pattern Recognition}, pages 8639--8648, 2018.

\bibitem{wang2020adversarial}
Qi Wang, Thierry Arti{\`e}res, Mickael Chen, and Ludovic Denoyer.
\newblock Adversarial learning for modeling human motion.
\newblock {\em The Visual Computer}, 36(1):141--160, 2020.

\bibitem{wang2020learning}
Zhenyi Wang, Ping Yu, Yang Zhao, Ruiyi Zhang, Yufan Zhou, Junsong Yuan, and
  Changyou Chen.
\newblock Learning diverse stochastic human-action generators by learning
  smooth latent transitions.
\newblock In {\em Proceedings of the AAAI conference on artificial
  intelligence}, volume~34, pages 12281--12288, 2020.

\bibitem{yan2019convolutional}
Sijie Yan, Zhizhong Li, Yuanjun Xiong, Huahan Yan, and Dahua Lin.
\newblock Convolutional sequence generation for skeleton-based action
  synthesis.
\newblock In {\em Proceedings of the IEEE/CVF International Conference on
  Computer Vision}, pages 4394--4402. IEEE Computer Society, 2019.

\bibitem{yan2018spatial}
Sijie Yan, Yuanjun Xiong, and Dahua Lin.
\newblock Spatial temporal graph convolutional networks for skeleton-based
  action recognition.
\newblock In {\em Thirty-second AAAI conference on artificial intelligence},
  2018.

\bibitem{yang2018action}
Zhengyuan Yang, Yuncheng Li, Jianchao Yang, and Jiebo Luo.
\newblock Action recognition with spatio--temporal visual attention on skeleton
  image sequences.
\newblock {\em IEEE Transactions on Circuits and Systems for Video Technology},
  29(8):2405--2415, 2018.

\bibitem{yu2020structure}
Ping Yu, Yang Zhao, Chunyuan Li, Junsong Yuan, and Changyou Chen.
\newblock Structure-aware human-action generation.
\newblock In {\em European Conference on Computer Vision}, pages 18--34.
  Springer, 2020.

\bibitem{yuan2020dlow}
Ye Yuan and Kris Kitani.
\newblock Dlow: Diversifying latent flows for diverse human motion prediction.
\newblock In {\em European Conference on Computer Vision}, pages 346--364.
  Springer, 2020.

\bibitem{zhang2021write}
Jia-Qi Zhang, Xiang Xu, Zhi-Meng Shen, Ze-Huan Huang, Yang Zhao, Yan-Pei Cao,
  Pengfei Wan, and Miao Wang.
\newblock Write-an-animation: High-level text-based animation editing with
  character-scene interaction.
\newblock In {\em Computer Graphics Forum}, volume~40, pages 217--228. Wiley
  Online Library, 2021.

\bibitem{zhang2021we}
Yan Zhang, Michael~J Black, and Siyu Tang.
\newblock We are more than our joints: Predicting how 3d bodies move.
\newblock In {\em Proceedings of the IEEE/CVF Conference on Computer Vision and
  Pattern Recognition}, pages 3372--3382, 2021.

\bibitem{zhou2018auto}
Yi Zhou, Zimo Li, Shuangjiu Xiao, Chong He, Zeng Huang, and Hao Li.
\newblock Auto-conditioned recurrent networks for extended complex human motion
  synthesis.
\newblock In {\em International Conference on Learning Representations}, 2018.

\end{thebibliography}
}

\end{document}